# Channel Coding: The Road to Channel Capacity

Daniel J. Costello, Jr., *Fellow, IEEE*, and G. David Forney, Jr., *Fellow, IEEE*

### Abstract

Starting from Shannon's celebrated 1948 channel coding theorem, we trace the evolution of channel coding from Hamming codes to capacity-approaching codes. We focus on the contributions that have led to the most significant improvements in performance vs. complexity for practical applications, particularly on the additive white Gaussian noise (AWGN) channel. We discuss algebraic block codes, and why they did not prove to be the way to get to the Shannon limit. We trace the antecedents of today's capacity-approaching codes: convolutional codes, concatenated codes, and other probabilistic coding schemes. Finally, we sketch some of the practical applications of these codes.

## I. INTRODUCTION

The field of channel coding started with Claude Shannon's 1948 landmark paper [1]. For the next half century, its central objective was to find practical coding schemes that could approach channel capacity (hereafter called "the Shannon limit") on well-understood channels such as the additive white Gaussian noise (AWGN) channel. This goal proved to be challenging, but not impossible. In the past decade, with the advent of turbo codes and the rebirth of low-density parity-check codes, it has finally been achieved, at least in many cases of practical interest.

As Bob McEliece observed in his 2004 Shannon Lecture [2], the extraordinary efforts that were required to achieve this objective may not be fully appreciated by future historians. McEliece imagined a biographical note in the 166th edition of the *Encyclopedia Galactica* along the following lines:

**Claude Shannon**: Born on the planet Earth (Sol III) in the year 1916 A.D. Generally regarded as the father of the Information Age, he formulated the notion of channel capacity in 1948 A.D. Within several decades, mathematicians and engineers had devised practical ways to communicate reliably at data rates within 1% of the Shannon limit ...

The purpose of this paper is to tell the story of how Shannon's challenge was met, at least as it appeared to us, before the details of this story are lost to memory.

We focus on the AWGN channel, which was the target for many of these efforts. In Section II, we review various definitions of the Shannon limit for this channel.

In Section III, we discuss the subfield of algebraic coding, which dominated the channel coding field for its first couple of decades. We will discuss both the achievements of algebraic coding, and also the reasons why it did not prove to be the way to approach the Shannon limit.

Daniel J. Costello, Jr., is with the Univ. Notre Dame, IN 46556, USA, e-mail: costello.2@nd.edu.

G. David Forney, Jr., is with the Mass. Inst. of Tech., Cambridge, MA 02139 USA, e-mail: forney@mit.edu.

This work was supported in part by NSF Grant CCR02-05310 and NASA Grant NNGO5GH73G.

In Section IV, we discuss the alternative line of development that was inspired more directly by Shannon's random coding approach, which is sometimes called "probabilistic coding." The first major contribution to this area after Shannon was Elias' invention of convolutional codes. This line of development includes product codes, concatenated codes, trellis decoding of block codes, and ultimately modern capacity-approaching codes.

In Section V, we discuss codes for bandwidth-limited channels, namely lattice codes and trellis-coded modulation.

Finally, in Section VI, we discuss the development of capacity-approaching codes, principally turbo codes and low-density parity-check (LDPC) codes.

## II. CODING FOR THE AWGN CHANNEL

A coding scheme for the AWGN channel may be characterized by two simple parameters: its signal-to-noise ratio (SNR) and its spectral efficiency $\eta$ in bits per second per Hertz (b/s/Hz). The SNR is the ratio of average signal power to average noise power, a dimensionless quantity. The spectral efficiency of a coding scheme that transmits $R$ bits per second (b/s) over an AWGN channel of bandwidth $W$ Hz is simply $\eta = R/W$ b/s/Hz.

Coding schemes for the AWGN channel typically map a sequence of bits at a rate $R$ b/s to a sequence of real symbols at a rate of $2B$ symbols per second; the discrete-time code rate is then $r = R/2B$ bits per symbol. The sequence of real symbols is then modulated via pulse amplitude modulation (PAM) or quadrature amplitude modulation (QAM) for transmission over an AWGN channel of bandwidth $W$. By Nyquist theory, $B$ (sometimes called the "Shannon bandwidth" [3]) cannot exceed the actual bandwidth $W$. If $B \approx W$, then the spectral efficiency is $\eta = R/W \approx R/B = 2r$. We therefore say that the *nominal spectral efficiency* of a discrete-time coding scheme is $2r$, the discrete-time code rate in bits per two symbols. The actual spectral efficiency $\eta = R/W$ of the corresponding continuous-time scheme is upperbounded by the nominal spectral efficiency $2r$, and approaches $2r$ as $B \rightarrow W$. Thus, for discrete-time codes, we will often denote $2r$ by $\eta$, implicitly assuming $B \approx W$.

Shannon showed that on an AWGN channel with signal-to-noise ratio SNR and bandwidth $W$ Hz, the rate of reliable transmission is upperbounded by

$$R < W \log_2(1 + \text{SNR}).$$

Moreover, if a long code with rate $R < W \log_2(1+\text{SNR})$ is chosen at random, then there exists a decoding scheme such that with high probability the code and decoder will achieve highly reliable transmission (*i.e.,* low probability of decoding error).

Equivalently, Shannon's result shows that the spectral efficiency is upperbounded by

$$\eta < \log_2(1 + \text{SNR});$$

or, given a spectral efficiency $\eta$, that the SNR needed for reliable transmission is lowerbounded by

$$\text{SNR} > 2^\eta - 1.$$

So we may say that the Shannon limit on rate (*i.e.,* the *channel capacity*) is $W \log_2(1 + \text{SNR})$ b/s, or equivalently that the Shannon limit on spectral efficiency is $\log_2(1 + \text{SNR})$ b/s/Hz, or equivalently that the Shannon limit on SNR for a given spectral efficiency $\eta$ is $2^\eta - 1$. Note that the Shannon limit on SNR is a lower bound rather than an upper bound.

These bounds suggest that we define a *normalized SNR* parameter $\text{SNR}_{\text{norm}}$ as follows:

$$\text{SNR}_{\text{norm}} = \frac{\text{SNR}}{2^\eta - 1}.$$

Then for any reliable coding scheme, $\text{SNR}_{\text{norm}} > 1$; *i.e.,* the Shannon limit (lower bound) on $\text{SNR}_{\text{norm}}$ is 1 (0 dB), independent of $\eta$. Moreover, $\text{SNR}_{\text{norm}}$ measures the "gap to capacity", *i.e.,* $10 \log_{10} \text{SNR}_{\text{norm}}$ is the difference in decibels (dB)[1] between the SNR actually used and the Shannon limit on SNR given $\eta$, namely $2^\eta - 1$. If the desired spectral efficiency is less than 1 b/s/Hz (the so-called *power-limited regime*), then it can be shown that binary codes can be used on the AWGN channel with a cost in Shannon limit on SNR of less than 0.2 dB. On the other hand, since for a binary coding scheme the discrete-time code rate is bounded by $r \leq 1$ bit per symbol, the spectral efficiency of a binary coding scheme is limited to $\eta \leq 2r \leq 2$ b/s/Hz, so multilevel coding schemes must be used if the desired spectral efficiency is greater than 2 b/s/Hz (the so-called *bandwidth-limited regime*). In practice, coding schemes for the power-limited and bandwidth-limited regimes differ considerably.

---

[1]In decibels, a multiplicative factor of $\alpha$ is expressed as $10 \log_{10} \alpha$ dB.

A closely related normalized SNR parameter that has been traditionally used in the power-limited regime is $E_b/N_0$, which may be defined as

$$E_b/N_0 = \frac{\text{SNR}}{\eta} = \frac{2^\eta - 1}{\eta}\text{SNR}_{\text{norm}}.$$

For a given spectral efficiency $\eta$, $E_b/N_0$ is thus lowerbounded by

$$E_b/N_0 > \frac{2^\eta - 1}{\eta},$$

so we may say that the Shannon limit (lower bound) on $E_b/N_0$ as a function of $\eta$ is $\frac{2^\eta - 1}{\eta}$. This function decreases monotonically with $\eta$, and approaches $\ln 2$ as $\eta \to 0$, so we may say that the *ultimate Shannon limit* (lower bound) on $E_b/N_0$ for any $\eta$ is $\ln 2$ (-1.59 dB).

We see that as $\eta \to 0$, $E_b/N_0 \to \text{SNR}_{\text{norm}} \ln 2$, so $E_b/N_0$ and $\text{SNR}_{\text{norm}}$ become equivalent parameters in the severely power-limited regime. In the power-limited regime, we will therefore use the traditional parameter $E_b/N_0$. However, in the bandwidth-limited regime, we will use $\text{SNR}_{\text{norm}}$, which is more informative in this regime.

## III. Algebraic coding

The algebraic coding paradigm dominated the first several decades of the field of channel coding. Indeed, most of the textbooks on coding of this period (including Peterson [4], Berlekamp [5], Lin [6], Peterson and Weldon [7], MacWilliams and Sloane [8], and Blahut [9]) covered only algebraic coding theory.

Algebraic coding theory is primarily concerned with linear $(n, k, d)$ block codes over the binary field $\mathbb{F}_2$. A binary linear $(n, k, d)$ block code consists of $2^k$ binary $n$-tuples, called codewords, which have the group property: *i.e.,* the componentwise mod-2 sum of any two codewords is another codeword. The parameter $d$ denotes the minimum Hamming distance between any two distinct codewords— *i.e.,* the minimum number of coordinates in which any two codewords differ. The theory generalizes to linear $(n, k, d)$ block codes over nonbinary fields $\mathbb{F}_q$.

The principal objective of algebraic coding theory is to maximize the minimum distance $d$ for a given $(n, k)$. The motivation for this objective is to maximize error-correction power. Over a binary symmetric channel (BSC: a binary-input, binary-output channel with statistically independent binary errors), the optimum decoding rule is to decode to the codeword closest in Hamming distance to the received $n$-tuple. With this rule, a code with minimum distance $d$ can correct all patterns of $(d-1)/2$ or fewer channel errors (assuming that $d$ is odd), but cannot correct some patterns containing a greater number of errors.

The field of algebraic coding theory has had many successes, which we will briefly survey below. However, even though binary algebraic block codes can be used on the AWGN channel, they have not proved to be the way to approach channel capacity on this channel, even in the power-limited regime. Indeed, they have not proved to be the way to approach channel capacity even on the BSC. As we proceed, we will discuss some of the fundamental reasons for this failure.

### A. Binary coding on the power-limited AWGN channel

A binary linear $(n, k, d)$ block code may be used on a Gaussian channel as follows.

To transmit a codeword, each of its $n$ binary symbols may be mapped into the two symbols $\{\pm\alpha\}$ of a binary pulse-amplitude-modulation (2-PAM) alphabet, yielding a two-valued real $n$-tuple $\mathbf{x}$. This $n$-tuple may then be sent through a channel of bandwidth $W$ at a symbol rate $2B$ up to the Nyquist limit of $2W$ binary symbols per second, using standard pulse amplitude modulation (PAM) for baseband channels, or quadrature amplitude modulation (QAM) for passband channels.

At the receiver, an optimum PAM or QAM detector can produce a real-valued $n$-tuple $\mathbf{y} = \mathbf{x} + \mathbf{n}$, where $\mathbf{x}$ is the transmitted sequence and $\mathbf{n}$ is a discrete-time white Gaussian noise sequence. The optimum (maximum likelihood) decision rule is then to choose the one of the $2^k$ possible transmitted sequences $\mathbf{x}$ that is closest to the received sequence $\mathbf{y}$ in Euclidean distance.

If the symbol rate $2B$ approaches the Nyquist limit of $2W$ symbols per second, then the transmitted data rate can approach $R = (k/n)2W$ b/s, so the spectral efficiency of such a binary coding scheme can approach $\eta = 2k/n$ b/s/Hz. As mentioned previously, since $k/n \leq 1$, we have $\eta \leq 2$ b/s/Hz; *i.e.,* binary coding cannot be used in the bandwidth-limited regime.

With no coding (independent transmission of random bits via PAM or QAM), the transmitted data rate is $2W$ b/s, so the nominal spectral efficency is $\eta = 2$ b/s/Hz. It is straightforward to show that with optimum modulation and detection the probability of error per bit is

$$P_b(E) = Q(\sqrt{\text{SNR}}) = Q(\sqrt{2E_b/N_0}),$$

where

$$Q(x) = \frac{1}{\sqrt{2\pi}} \int_x^\infty e^{-y^2/2} dy$$

is the Gaussian probability of error function. This baseline performance curve of $P_b(E)$ *vs.* $E_b/N_0$ for uncoded transmission is plotted in Figure 1. For example, in order to achieve a bit error probability of $P_b(E) \approx 10^{-5}$, we must have $E_b/N_0 \approx 9.1$ (9.6 dB) for uncoded transmission.

On the other hand, the Shannon limit on $E_b/N_0$ for $\eta = 2$ is $E_b/N_0 = 1.5$ (1.76 dB), so the gap to capacity at the uncoded binary-PAM spectral efficiency of $\eta = 2$ is $\text{SNR}_{\text{norm}} \approx 7.8$ dB. If a coding scheme with unlimited bandwidth expansion were allowed, *i.e.,* $\eta \to 0$, then a further gain of 3.35 dB to the ultimate Shannon limit on $E_b/N_0$ of -1.59 dB would be achievable. These two limits are also shown on Figure 1.

The performance curve of any practical coding scheme that improves on uncoded transmission must lie between the relevant Shannon limit and the uncoded performance curve. Thus Figure 1 defines the "playing field" for channel coding. The *real coding gain* of a coding scheme at a given probability of error per bit $P_b(E)$ will be defined as the difference (in dB) between the $E_b/N_0$ required to obtain that $P_b(E)$ with coding *vs.* without coding. Thus the maximum possible real coding gain at $P_b(E) \approx 10^{-5}$ is about 11.2 dB.

For moderate-complexity binary linear $(n, k, d)$ codes, it can often be assumed that the decoding error probability is dominated by the probability of making an error to one of the nearest-neighbor codewords. If this assumption holds, then it is easy to show that with optimum (minimum-Euclidean-distance) decoding, the decoding error probability $P_B(E)^2$ per block is well approximated by the union bound estimate

$$P_B(E) \approx N_d Q(\sqrt{d \cdot \text{SNR}}) = N_d Q(\sqrt{(dk/n)2E_b/N_0}),$$

where $N_d$ denotes the number of codewords of Hamming weight $d$. The probability of decoding error per information bit $P_b(E)$ is then given by

$$P_b(E) = P_B(E)/k \approx (N_d/k) \ Q(\sqrt{(dk/n)2E_b/N_0}) = (N_d/k) \ Q(\sqrt{\gamma_c 2E_b/N_0}),$$

where the quantity $\gamma_c = dk/n$ is called the *nominal coding gain* of the code.[3] The real coding gain is less than the nominal coding gain $\gamma_c$ if the "error coefficient" $N_d/k$ is greater than 1. A rule of thumb that is valid when the error coefficient $N_d/k$ is not too large and $P_b(E)$ is on the order of $10^{-6}$ is that a factor of 2 increase in the error coefficient costs about 0.2 dB of real coding gain. As $P_b(E) \to 0$, the real coding gain approaches the nominal coding gain $\gamma_c$, so $\gamma_c$ is also called the *asymptotic coding gain*.

---

[2]The probability of error per information bit is not in general the same as the bit error probability (average number of bit errors per transmitted bit), although both normally have the same exponent (argument of the $Q$ function).

[3]An $(n, k, d)$ code with odd minimum distance $d$ may be extended by addition of an overall parity-check to an $(n + 1, k, d + 1)$ even-minimum-distance code. For error correction, such an extension is of no use, since the extended code corrects no more errors but has a lower code rate; but for an AWGN channel, such an extension always helps (unless $k = 1$ and $d = n$), since the nominal coding gain $\gamma_c = dk/n$ increases. Thus an author who discusses odd-distance codes is probably thinking about minimum-Hamming-distance decoding, whereas an author who discusses even-distance codes is probably thinking about minimum-Euclidean-distance decoding.

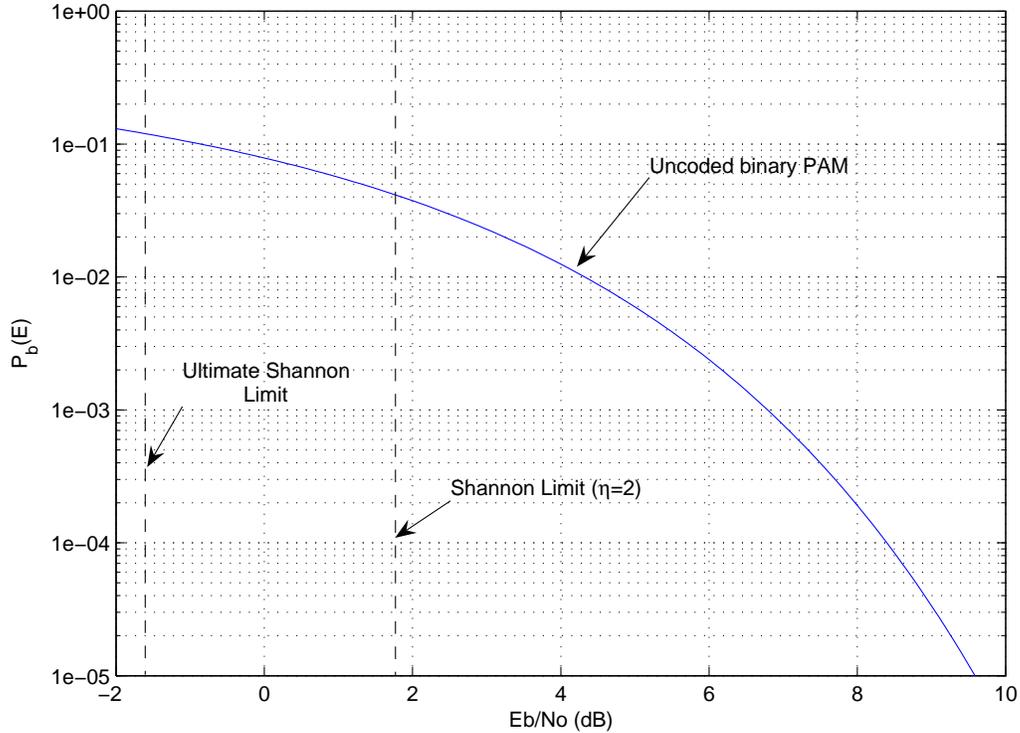

Fig. 1. $P_b(E)$ vs. $E_b/N_0$ for uncoded binary PAM, compared to Shannon limits on $E_b/N_0$ for $\eta = 2$ and $\eta \to 0$.

For example, consider the binary linear $(32, 6, 16)$ "biorthogonal" block code, so called because the Euclidean images of the 64 codewords consist of 32 orthogonal vectors and their negatives. With this code, every codeword has $N_d = 62$ nearest neighbors at minimum Hamming distance $d = 16$. Its nominal spectral efficiency is $\eta = 3/8$, its nominal coding gain is $\gamma_c = 3$ (4.77 dB), and its probability of decoding error per information bit is

$$P_b(E) \approx (62/6) \; Q(\sqrt{6E_b/N_0}),$$

which is plotted in Figure 2. We see that this code requires $E_b/N_0 \approx 5.8$ dB to achieve $P_b(E) \approx 10^{-5}$, so its real coding gain at this error probability is about 3.8 dB.

At this point, we can already identify two issues that must be addressed to approach the Shannon limit on AWGN channels. First, in order to obtain optimum performance, the decoder must operate on the real-valued received sequence $\mathbf{y}$ ("soft decisions") and minimize Euclidean distance, rather than quantize to a two-level received sequence ("hard decisions") and minimize Hamming distance. It can be shown that hard decisions (two-level quantization) generally costs 2 to 3 dB in decoding performance. Thus, in order to approach the Shannon limit on an AWGN channel, the error-correction paradigm of algebraic coding must be modified to accommodate soft decisions.

Second, we can see already that decoding complexity is going to be an issue. For optimum decoding, soft decision or hard, the decoder must choose the best of $2^k$ codewords, so a straightforward exhaustive optimum decoding algorithm will require on the order of $2^k$ computations. Thus, as codes become large, lower-complexity decoding algorithms that approach optimum performance must be devised.

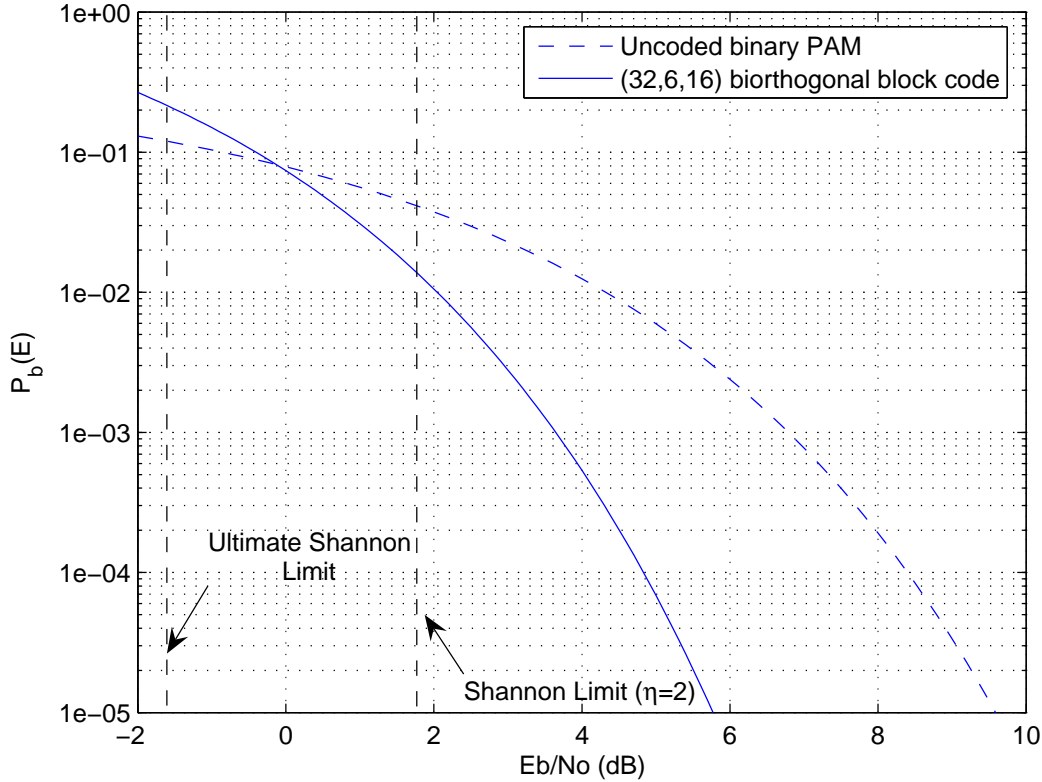

Fig. 2. $P_b(E)$ *vs.* $E_b/N_0$ for the $(32, 6, 16)$ biorthogonal block code, compared to uncoded PAM and Shannon limits.

### B. The earliest codes: Hamming, Golay, and Reed-Muller

The first nontrivial code to appear in the literature was the $(7, 4, 3)$ Hamming code, mentioned by Shannon in his original paper [1]. Richard Hamming, a colleague of Shannon at Bell Labs, developed an infinite class of single-error-correcting ($d = 3$) binary linear codes, with parameters ($n = 2^m - 1, k = 2^m - 1 - m, d = 3$) for $m \geq 2$ [10]. Thus $k/n \to 1$ and $\eta \to 2$ as $m \to \infty$, while $\gamma_c \to 3$ (4.77 dB). However, even with optimum soft-decision decoding, the real coding gain of Hamming codes on the AWGN channel never exceeds about 3 dB.

The Hamming codes are "perfect," in the sense that the spheres of Hamming radius 1 about each of the $2^k$ codewords contain $2^m$ binary $n$-tuples and thus form a "perfect" (exhaustive) partition of binary $n$-space $(\mathbb{F}_2)^n$. Shortly after the publication of Shannon's paper, the Swiss mathematician Marcel Golay published a half-page paper [11] with a "perfect" binary linear $(23, 12, 7)$ triple-error-correcting code, in which the spheres of Hamming radius 3 about each of the $2^{12}$ codewords (containing $\binom{23}{0} + \binom{23}{1} + \binom{23}{2} + \binom{23}{3} = 2^{11}$ binary $n$-tuples) form an exhaustive partition of $(\mathbb{F}_2)^{23}$— and also a similar "perfect" $(11, 6, 5)$ double-error-correcting ternary code. These binary and ternary Golay codes have come to be considered probably the most remarkable of all algebraic block codes, and it is now known that no other nontrivial "perfect" linear codes exist. Berlekamp [12] characterized Golay's paper as the "best single published page" in coding theory during 1948–1973.

Another early class of error-correcting codes was the Reed-Muller (RM) codes, which were introduced in 1954 by David Muller [13], and then reintroduced shortly thereafter with an efficient decoding algorithm by Irving Reed [14]. The RM$(r, m)$ codes are a class of multiple-error-correcting $(n, k, d)$ codes parametrized by two integers $r$ and $m$, $0 \leq r \leq m$, such that $n = 2^m$ and $d = 2^{m-r}$. The RM$(0, m)$ code is the $(2^m, 1, 2^m)$ binary repetition code (consisting of two codewords, the all-zero and all-one words), and the RM$(m, m)$ code is the $(2^m, 2^m, 1)$ binary code consisting of all binary $2^m$-tuples (*i.e.,* uncoded transmission).

Starting with RM$(0, 1) = (2, 1, 2)$ and RM$(1, 1) = (2, 2, 1)$, the RM codes may be constructed recursively by the length-doubling $|u|u+v|$ (Plotkin, squaring) construction as follows:

$$\text{RM}(r,m) = \{(\mathbf{u}, \mathbf{u} + \mathbf{v}) \mid \mathbf{u} \in \text{RM}(r, m-1), \mathbf{v} \in \text{RM}(r-1, m-1)\}.$$

From this construction it follows that the dimension $k$ of $\text{RM}(r,m)$ is given recursively by

$$k(r,m) = k(r, m-1) + k(r-1, m-1),$$

or nonrecursively by $k(r,m) = \sum_{i=0}^{r} \binom{m}{i}$.

Figure 3 shows the parameters of the RM codes of lengths $\leq 32$ in a tableau that reflects this length-doubling construction. For example, the $\text{RM}(2,5)$ code is a $(32, 16, 8)$ code that can be constructed from the $\text{RM}(2,4) = (16, 11, 4)$ code and the $\text{RM}(1,4) = (16, 5, 8)$ code.

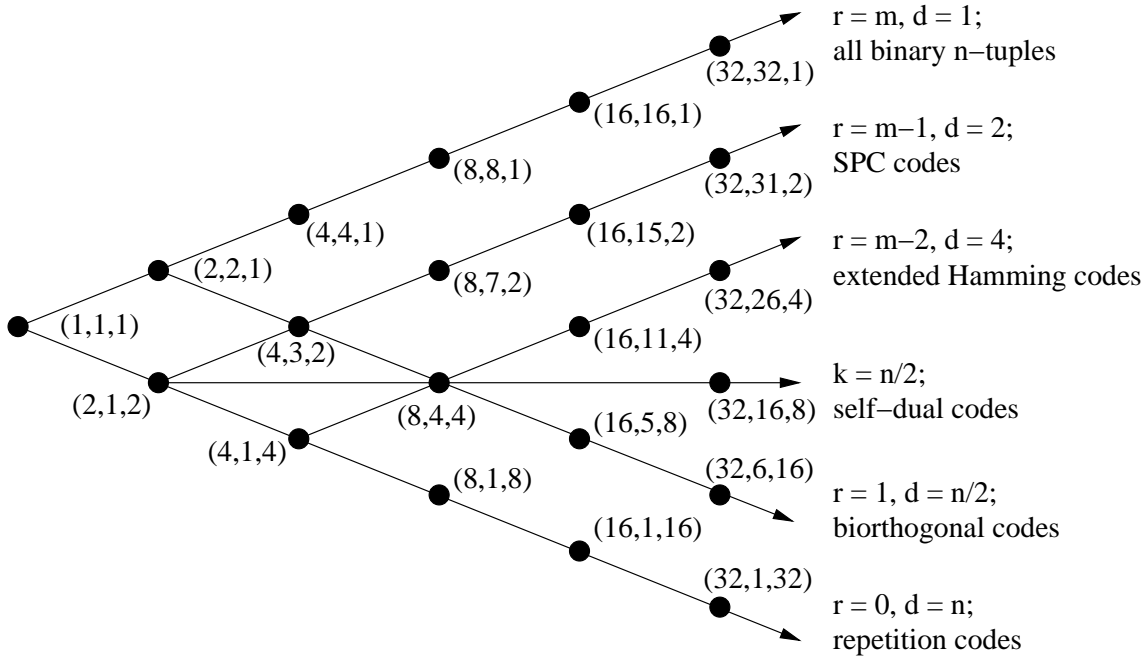

Fig. 3.   Tableau of Reed-Muller codes.

RM codes include several important subclasses of codes. We have already mentioned the $(2^m, 2^m, 1)$ codes consisting of all binary $2^m$-tuples and the $(2^m, 1, 2^m)$ repetition codes. RM codes also include the $(2^m, 2^m - 1, 2)$ single-parity-check (SPC) codes, the $(2^m, 2^m - m - 1, 4)$ extended Hamming[4] codes, the $(2^m, m+1, 2^{m-1})$ biorthogonal codes, and, for odd $m$, a class of $(2^m, 2^{m-1}, 2^{(m+1)/2})$ self-dual codes.[5]

Reed [14] introduced a low-complexity hard-decision error-correction algorithm for RM codes based on a simple *majority-logic decoding rule*. This simple decoding rule is able to correct all hard-decision error patterns of weight $\lfloor (d-1)/2 \rfloor$ or less, which is the maximum possible (*i.e.,* it is a *bounded-distance decoding rule*). This simple majority-logic, hard-decision decoder was attractive for the technology of the 50s and 60s.

RM codes are thus an infinite class of codes with flexible parameters that can achieve near-optimal decoding on a BSC with a simple decoding algorithm. This was an important advance over the Hamming and Golay codes, whose parameters are much more restrictive.

Performance curves with optimum hard-decision coding are shown in Figure 4 for the $(31, 26, 3)$ Hamming code, the $(23, 12, 7)$ Golay code, and the $(31, 16, 7)$ shortened RM code. We see that they achieve real coding gains at $P_b(E) \approx 10^{-5}$ of only 0.9 dB, 2.3 dB, and 1.6 dB, respectively. The reasons for this poor performance are the use of hard decisions, which costs roughly 2 dB, and the fact that by modern standards these codes are very short.

---

[4]An $(n, k, d)$ code can be extended by adding code symbols or shortened by deleting code symbols; see footnote 3.

[5]An $(n, k, d)$ binary linear code forms a $k$-dimensional subspace of the vector space $\mathbb{F}_2^n$. The dual of an $(n, k, d)$ code is the $(n-k)$-dimensional orthogonal subspace. A code that equals its dual is called self-dual. For self-dual codes, it follows that $k = n/2$.

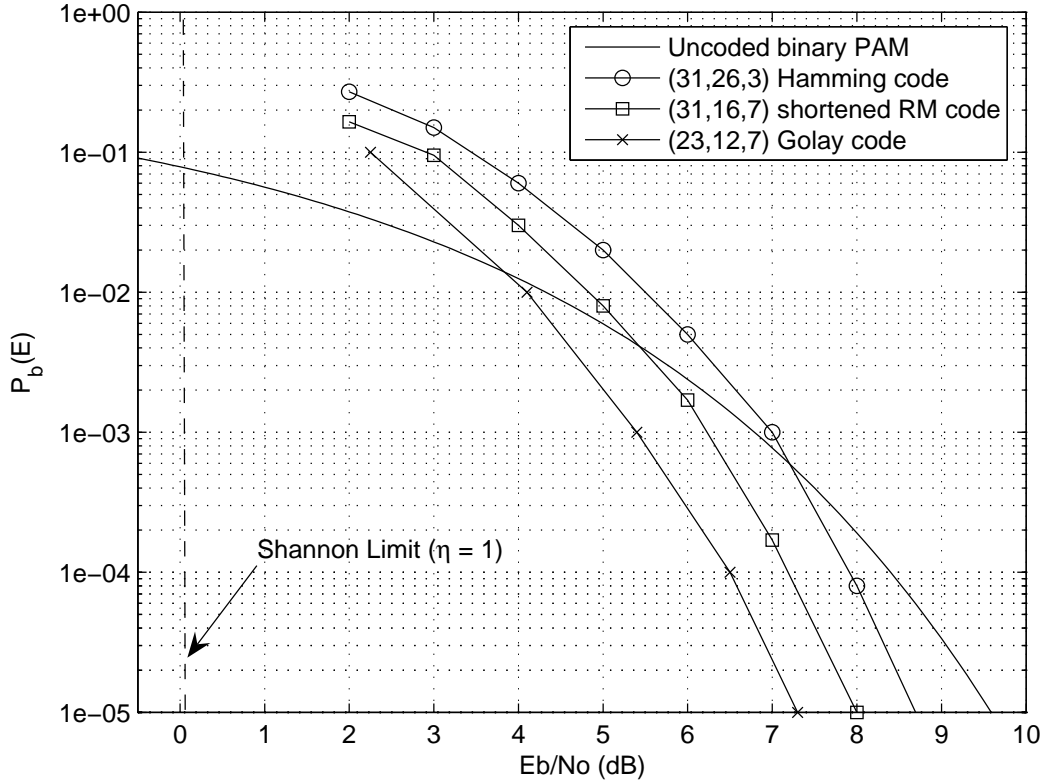

Fig. 4. $P_b(E)$ vs. $E_b/N_0$ for the (31,26,3) Hamming code, the (23,12,7) Golay code, and the (31,16,7) shortened RM code with optimum hard-decision decoding, compared to uncoded binary-PAM.

It is clear from the tableau of Figure 3 that RM codes are not asymptotically "good"; that is, there is no sequence of $(n, k, d)$ RM codes of increasing length $n$ such that both $k/n$ and $d/n$ are bounded away from 0 as $n \to \infty$. Since asymptotic goodness was the Holy Grail of algebraic coding theory (it is easy to show that typical random binary codes are asymptotically good), and since codes with somewhat better $(n, k, d)$ (e.g., BCH codes) were found subsequently, theoretical attention soon turned away from RM codes.

However, in recent years it has been recognized that "RM codes are not so bad." RM codes are particularly good in terms of performance vs. complexity with trellis-based decoding and other soft-decision decoding algorithms, as we note in Section IV-E. Finally, they are almost as good in terms of $(n, k, d)$ as the best binary codes known for lengths less than 128, which is the principal application domain of algebraic block codes.

Indeed, with optimum decoding, RM codes may be "good enough" to reach the Shannon limit on the AWGN channel. Notice that the nominal coding gains of the self-dual RM codes and the biorthogonal codes become infinite as $m \to \infty$. It is known that with optimum (minimum-Euclidean-distance) decoding, the real coding gain of the biorthogonal codes does asymptotically approach the ultimate Shannon limit, albeit with exponentially increasing complexity and vanishing spectral efficiency. It seems likely that the real coding gains of the self-dual RM codes with optimum decoding approach the Shannon limit at the nonzero spectral efficiency of $\eta = 1$, albeit with exponential complexity, but to our knowledge this has never been proved.

## C. Soft decisions: Wagner decoding

On the road to modern capacity-approaching codes for AWGN channels, an essential step has been to replace hard-decision with soft-decision decoding; i.e., decoding that takes into account the reliability of received channel outputs.

The earliest soft-decision decoding algorithm known to us is *Wagner decoding*, described in [15] and attributed to C. A. Wagner, which is an optimum decoding rule for the special class of $(n, n-1, 2)$ single-parity-check (SPC) codes. Each received real-valued symbol $r_k$ from an AWGN channel may be represented in sign-magnitude form, where the sign $\mathrm{sgn}(r_k)$ indicates the "hard decision," and the magnitude $|r_k|$ indicates the "reliability" of $r_k$. The Wagner rule is: first check whether the hard-decision binary $n$-tuple is a codeword. If so, accept it. If not, then flip the hard decision corresponding to the output $r_k$ that has the minimum reliability $|r_k|$.

It is easy to show that the Wagner rule finds the minimum-Euclidean-distance codeword; *i.e.,* that Wagner decoding is optimum for an $(n, n-1, 2)$ SPC code. Moreover, Wagner decoding is much simpler than exhaustive minimum-distance decoding, which requires on the order of $2^{n-1}$ computations.

## D. BCH and Reed-Solomon codes

In the 1960s, research in channel coding was dominated by the development of algebraic block codes, particularly cyclic codes. The algebraic coding paradigm used the structure of finite-field algebra to design efficient encoding and error-correction procedures for linear block codes operating on a hard-decision channel. The emphasis was on constructing codes with a guaranteed minimum distance $d$, and then using the algebraic structure of the codes to design bounded-distance error-correction algorithms whose complexity grows only as a small power of $d$. In particular, the goal was to develop flexible classes of easily-implementable codes with better performance than RM codes.

Cyclic codes are codes that are invariant under cyclic ("end-around") shifts of $n$-tuple codewords. They were first investigated by Eugene Prange in 1957 [16], and became the primary focus of research after the publication of Wesley Peterson's pioneering text in 1961 [4]. Cyclic codes have a nice algebraic theory, and attractively simple encoding and decoding procedures based on cyclic shift-register implementations. Hamming, Golay, and shortened RM codes can be put into cyclic form.

The "big bang" in this field was the invention of Bose-Chaudhuri-Hocquenghem (BCH) and Reed-Solomon (RS) codes in three independent papers in 1959 and 1960 [17], [18], [19]. It was shortly recognized that RS codes are a class of nonbinary BCH codes, or alternatively that BCH codes are subfield subcodes of RS codes.

Binary BCH codes include a large class of $t$-error-correcting cyclic codes of length $n = 2^m - 1$, odd minimum distance $d = 2t + 1$, and dimension $k \geq n - mt$. Compared to shortened RM codes of a given length $n = 2^m - 1$, there are more codes from which to choose, and for $n \geq 63$ the BCH codes can have a somewhat larger dimension $k$ for a given minimum distance $d$. However, BCH codes are still not asymptotically "good." Although they are the premier class of binary algebraic block codes, they have not been used much in practice, except as "cyclic redundancy check" (CRC) codes for error detection in automatic-repeat-request (ARQ) systems.

In contrast, the nonbinary Reed-Solomon codes have proved to be highly useful in practice (although not necessarily in cyclic form). An (extended or shortened) RS code over the finite field $\mathbb{F}_q, q = 2^m$, can have any block length up to $n = q + 1$, any minimum distance $d \leq n$ (where Hamming distance is defined in terms of $q$-ary symbols), and dimension $k = n - d + 1$, which meets an elementary upper bound called the Singleton bound [20]. In this sense, RS codes are optimum.

An important property of RS and BCH codes is that they can be efficiently decoded by algebraic decoding algorithms using finite-field arithmetic. A glance at the tables of contents of the IEEE TRANSACTIONS ON INFORMATION THEORY shows that the development of such algorithms was one of the most active research fields of the 1960s.

Already by 1960, Peterson had developed an error-correction algorithm with complexity on the order of $d^3$ [21]. In 1968, Elwyn Berlekamp [5] devised an error-correction algorithm with complexity on the order of $d^2$, which was interpreted by Jim Massey [22] as an algorithm for finding the shortest linear feedback shift register that can generate a certain sequence. This *Berlekamp-Massey algorithm* became the standard for the next decade. Finally, it was shown that these algorithms could be straightforwardly extended to correct both erasures and errors [23], and even to correct soft decisions [24], [25] (suboptimally, but in some cases asymptotically optimally).

The fact that RS codes are inherently nonbinary (the longest binary RS code has length 3) may cause difficulties in using them over binary channels. If the $2^m$-ary RS code symbols are simply represented as binary $m$-tuples and sent over a binary channel, then a single binary error can cause an entire $2^m$-ary symbol to be incorrect; this causes RS codes to be inferior to BCH codes as binary-error-correcting codes. However, in this mode RS codes are inherently good burst-error-correcting codes, since the effect of an $m$-bit burst that is concentrated in a single RS code symbol is only a single symbol error. In fact, it can be shown that RS codes are effectively optimal binary burst-error-correcting codes [26].

The ability of RS codes to correct both random and burst errors makes them particularly well suited for applications such as magnetic tape and disk storage, where imperfections in the storage media sometimes cause bursty errors. They are also useful as outer codes in concatenated coding schemes, to be discussed in Section IV-D. For these reasons, RS codes are probably the most widely deployed codes in practice.

### E. Reed-Solomon code implementations

The first major application of RS codes was as outer codes in concatenated coding systems for deep-space communications. For the 1977 Voyager mission, the Jet Propulsion Laboratory (JPL) used a $(255, 223, 33)$, 16-error-correcting RS code over $\mathbb{F}_{256}$ as an outer code, with a rate-1/2, 64-state convolutional inner code (see also Section IV-D). The RS decoder used special-purpose hardware for decoding, and was capable of running up to about 1 Mb/s [27]. This concatenated convolutional/RS coding system became a NASA standard.

1980 saw the first major commercial application of RS codes in the compact disc (CD) standard. This system used two short RS codes over $\mathbb{F}_{256}$, namely $(32, 28, 5)$ and $(28, 24, 5)$ RS codes, and operated at bit rates of the order of 4 Mb/s [28]. All subsequent audio and video magnetic storage systems have used RS codes for error correction, nowadays at much higher rates.

Cyclotomics, Inc. built a prototype "hypersystolic" RS decoder in 1986–88 that was capable of decoding a $(63, 53, 11)$ RS code over $\mathbb{F}_{64}$ at bit rates approaching 1 Gb/s [29]. This decoder may still hold the RS decoding speed record.

Reed-Solomon codes continue to be preferred for error correction when the raw channel error rate is not too large, because they can provide substantial error-correction power with relatively small redundancy at data rates up to tens or hundreds of Mb/s. They also work well against bursty errors. In these respects, they complement modern capacity-approaching codes.

### F. The "coding is dead" workshop

The first IEEE Communication Theory Workshop in St. Petersburg, Florida in April 1971 became famous as the "coding is dead" workshop. No written record of this workshop seems to have survived. However, Bob Lucky wrote a column about it many years later in IEEE SPECTRUM [30]. Lucky recalls:

A small group of us in the communications field will always remember a workshop held in Florida about 20 years ago ... One of my friends [Ned Weldon] gave a talk that has lived in infamy as the "coding is dead" talk. His thesis was that he and the other coding theorists formed a small, inbred group that had been isolated from reality for too long. He illustrated this talk with a single slide showing a pen of rats that psychologists had penned in a confined space for an extensive period of time. I cannot tell you what those rats were doing, but suffice it to say that the slide has since been borrowed many times to depict the depths of depravity into which such a disconnected group can fall ...

Of course, as Lucky goes on to say, the irony is that since 1971 coding has flourished and become embedded in practically all communications applications. He asks plaintively, "Why are we technologists so bad at predicting the future of technology?"

From today's perspective, one answer to this question could be that what Weldon was really asserting was that "algebraic coding is dead" (or at least had reached the point of diminishing returns).

Another answer was given on the spot by Irwin Jacobs, who stood up in the back row, flourished a medium-scale-integrated circuit (perhaps a 4-bit shift register), and asserted that "This is the future of coding." Elwyn Berlekamp said much the same thing. Interestingly, Jacobs and Berlekamp went on to lead the two principal coding companies of the 1970s, Linkabit and Cyclotomics, the one championing convolutional codes, and the other, block codes.

History has shown that both answers were right. Coding has moved from theory to practice in the past 35 years because (a) other classes of coding schemes have supplanted the algebraic coding paradigm, and (b) advances in integrated circuit technology have ultimately allowed designers to implement any (polynomial-complexity) algorithm that they can think of. Today's technology is on the order of a million times faster than that of 1971. Even though Moore's Law had already been propounded in 1971, it seems to be hard for the human mind to grasp what a factor of $10^6$ can make possible.

### G. Further developments in algebraic coding theory

Of course algebraic coding theory has not died; it continues to be an active research area. A recent text in this area is Roth [31].

A new class of block codes based on algebraic geometry (AG) was introduced by Goppa in the late 1970s [32], [33]. Tsfasman, Vladut, and Zink [34] constructed AG codes over nonbinary fields $\mathbb{F}_q$ with $q \geq 49$ whose minimum distance as $n \to \infty$ surpasses the Gilbert-Varshamov bound (the best known lower bound on the minimum distance of block codes), which is perhaps the most notable achievement of AG codes. AG codes are generally much longer than RS codes, and can usually be decoded by extensions of RS decoding algorithms. However, AG codes have not been adopted yet for practical applications. For a nice survey of this field, see [35].

In 1997, Sudan [36] introduced a list decoding algorithm based on polynomial interpolation for decoding beyond the guaranteed error-correction distance of RS and related codes.[6] Although in principle there may be more than one codeword within such an expanded distance, in fact with high probability only one will occur. Guruswami and Sudan [38] further improved the algorithm and its decoding radius, and Koetter and Vardy [39] extended it to handle soft decisions. There is currently some hope that algorithms of this type will be used in practice.

Other approaches to soft-decision decoding algorithms have continued to be developed, notably the ordered-statistics approach of Fossorier and Lin (see, *e.g.,* [40]) whose roots can be traced back to Wagner decoding.

### IV. PROBABILISTIC CODING

"Probabilistic coding" is a name for an alternative line of development that was more directly inspired by Shannon's probabilistic approach to coding. Whereas algebraic coding theory aims to find specific codes that maximize the minimum distance $d$ for a given $(n, k)$, probabilistic coding is more concerned with finding classes of codes that optimize average performance as a function of coding and decoding complexity. Probabilistic decoders typically use soft-decision (reliability) information, both as inputs (from the channel outputs), and at intermediate stages of the decoding process. Classical coding schemes that fall into this class include convolutional codes, product codes, concatenated codes, trellis-coded modulation, and trellis decoding of block codes. Popular textbooks that emphasize the probabilistic view of coding include Wozencraft and Jacobs [41], Gallager [42], Clark and Cain [43], Lin and Costello [44], Johannesson and Zigangirov [45], and the forthcoming book by Richardson and Urbanke [46].

For many years, the competition between the algebraic and probabilistic approaches was cast as a competition between block codes and convolutional codes. Convolutional coding was motivated from the start by the objective of optimizing the tradeoff of performance *vs.* complexity, which on the binary-input AWGN channel necessarily implies soft decisions and quasi-optimal decoding. In practice, most channel coding systems have used convolutional codes. Modern capacity-approaching codes are the ultimate fruit of this line of development.

---

[6]List decoding was an (unpublished) invention of Elias— see [37].

## A. Elias' invention of convolutional codes

Convolutional codes were invented by Peter Elias in 1955 [47]. Elias' goal was to find classes of codes for the binary symmetric channel (BSC) with as much structure as possible, without loss of performance.

Elias' several contributions have been nicely summarized by Bob Gallager, who was Elias' student [48]:

> [Elias'] 1955 paper . . . was perhaps the most influential early paper in information theory after Shannon's. This paper takes several important steps toward realizing the promise held out by Shannon's paper . . . .
>
> The first major result of the paper is a derivation of upper and lower bounds on the smallest achievable error probability on a BSC using codes of a given block length $n$. These bounds decrease exponentially with $n$ for any data rate $R$ less than the capacity $C$. Moreover, the upper and lower bounds are substantially the same over a significant range of rates up to capacity. This result shows that:
>
> (a) achieving a small error probability at any error rate near capacity necessarily requires a code with a long block length; and
>
> (b) almost all randomly chosen codes perform essentially as well as the best codes; that is, most codes are good codes.
>
> Consequently, Elias turned his attention to finding classes of codes that have some special structure, so as to simplify implementation, without sacrificing average performance over the class.
>
> His second major result is that the special class of linear codes has the same average performance as the class of completely random codes. Encoding of linear codes is fairly simple, and the symmetry and special structure of these codes led to a promise of simplified decoding strategies . . . . In practice, practically all codes are linear.
>
> Elias' third major result was the invention of [linear time-varying] convolutional codes . . . . These codes are even simpler to encode than general linear codes, and they have many other useful qualities. Elias showed that convolutional codes also have the same average performance as randomly chosen codes.

We may mention at this point that Gallager's doctoral thesis on low-density parity-check (LDPC) codes, supervised by Elias, was similarly motivated by the problem of finding a class of "random-like" codes that could be decoded near capacity with quasi-optimal performance and feasible complexity [49].

Linearity is the only algebraic property that is shared by convolutional codes and algebraic block codes.[7] The additional structure introduced by Elias was later understood as the dynamical structure of a discrete-time, $k$-input, $n$-output finite-state Markov process. A convolutional code is characterized by its code rate $k/n$, where $k$ and $n$ are typically small integers, and by the number of its states, which is often closely related to decoding complexity.

In more recent terms, Elias' and Gallager's codes can be represented as "codes on graphs," in which the complexity of the graph increases only linearly with the code block length. This is why convolutional codes are useful as components of turbo coding systems. In this light, there is a fairly straight line of development from Elias' invention to modern capacity-approaching codes. Nonetheless, this development actually took the better part of a half-century.

## B. Convolutional codes in the 1960s and 1970s

Shortly after Elias' paper, Jack Wozencraft recognized that the tree structure of convolutional codes permits decoding by a sequential search algorithm [51]. Sequential decoding became the subject of intense research at MIT, culminating in the development of the fast, storage-free Fano sequential decoding algorithm [52], and an analytical proof that the rate of a sequential decoding system is bounded by the computational cut-off rate $R_0$ [53].

Subsequently, Jim Massey proposed a very simple decoding method for convolutional codes, called threshold decoding [54]. Burst-error-correcting variants of threshold decoding developed by Massey and Gallager proved to be quite suitable for practical error correction [26]. Codex Corp. was founded in 1962 around the Massey and Gallager codes (including LDPC codes, which were never seriously considered for practical implementation). Codex built hundreds of burst-error-correcting threshold decoders during the 1960s, but the business never grew very large, and Codex left it in 1970.

---

[7]Linear convolutional codes have the algebraic structure of discrete-time multi-input, multi-output linear dynamical systems [50], but this is rather different from the algebraic structure of linear block codes.

In 1967, Andy Viterbi introduced what became known as the Viterbi algorithm (VA) as an "asymptotically optimal" decoding algorithm for convolutional codes, in order to prove exponential error bounds [55]. It was quickly recognized [56], [57] that the VA was actually an optimum decoding algorithm. More importantly, Jerry Heller at the Jet Propulsion Laboratory (JPL) [58], [59] realized that relatively short convolutional codes decoded by the VA were potentially quite practical— *e.g.*, a 64-state code could obtain a sizable real coding gain, on the order of 6 dB.

Linkabit Corp. was founded by Irwin Jacobs, Len Kleinrock, and Andy Viterbi in 1968 as a consulting company. In 1969, Jerry Heller was hired as Linkabit's first full-time employee. Shortly thereafter, Linkabit built a prototype 64-state Viterbi algorithm decoder ("a big monster filling a rack" [60]), capable of running at 2 Mb/s [61].

During the 1970s, through the leadership of Linkabit and JPL, the VA became part of the NASA standard for deep-space communication. Around 1975, Linkabit developed a relatively inexpensive, flexible, and fast VA chip. The VA soon began to be incorporated into many other communications applications.

Meanwhile, although a convolutional code with sequential decoding was the first code in space (for the 1968 Pioneer 9 mission [56]), and a few prototype sequential decoding systems were built, sequential decoding never took off in practice. By the time electronics technology could support sequential decoding, the VA had become a more attractive alternative. However, there seems to be a current resurgence of interest in sequential decoding for specialized applications [62].

### C. Soft decisions: APP decoding

Part of the attraction of convolutional codes is that all of these convolutional decoding algorithms are inherently capable of using soft decisions, without any essential increase in complexity. In particular, the VA implements minimum-Euclidean-distance sequence detection on an AWGN channel.

An alternative approach to using reliability information is to try to compute (exactly or approximately) the *a posteriori* probability (APP) of each transmitted bit being a 0 or a 1, given the APPs of each received symbol. In his thesis, Gallager [49] developed an iterative message-passing APP decoding algorithm for LDPC codes, which seems to have been the first appearance in any literature of the now-ubiquitous "sum-product algorithm" (also called "belief propagation"). At about the same time, Massey [54] developed an APP version of threshold decoding.

In 1974, Bahl, Cocke, Jelinek, and Raviv [63] published an algorithm for APP decoding of convolutional codes, now called the BCJR algorithm. Because this algorithm is more complicated than the VA (for one thing, it is a forward-backward rather than a forward-only algorithm) and its performance is more or less the same, it did not supplant the VA for decoding convolutional codes. However, because it is a soft-input, soft-output (SISO) algorithm (*i.e.*, APPs in, APPs out), it became a key element of iterative turbo decoding (see Section VI). Theoretically, it is now recognized as an implementation of the sum-product algorithm on a trellis.

### D. Product codes and concatenated codes

Before inventing convolutional codes, Elias had invented another class of codes now known as *product codes* [64]. The product of an $(n_1, k_1, d_1)$ with an $(n_2, k_2, d_2)$ binary linear block code is an $(n_1 n_2, k_1 k_2, d_1 d_2)$ binary linear block code. A product code may be decoded, simply but suboptimally, by independent decoding of the component codes. Elias showed that with a repeated product of extended Hamming codes, an arbitrarily low error probability could be achieved at a nonzero code rate, albeit at a code rate well below the Shannon limit.

In 1966, Dave Forney introduced *concatenated codes* [65]. As originally conceived, a concatenated code involves a serial cascade of two linear block codes: an *outer* $(n_2, k_2, d_2)$ nonbinary Reed-Solomon code over a finite field $\mathbb{F}_q$ with $q = 2^{k_1}$ elements, and an *inner* $(n_1, k_1, d_1)$ binary code with $q = 2^{k_1}$ codewords (see Figure 5). The resulting concatenated code is an $(n_1 n_2, k_1 k_2, d_1 d_2)$ binary linear block code. The key idea is that the inner and outer codes may be relatively short codes that are easy to encode and decode, whereas the concatenated code is a longer, more powerful code. For example, if the outer code is a $(15, 11, 5)$ RS code over $\mathbb{F}_{16}$ and the inner code is a $(7, 4, 3)$ binary Hamming code, then the concatenated code is a much more powerful $(105, 44, 15)$ code.

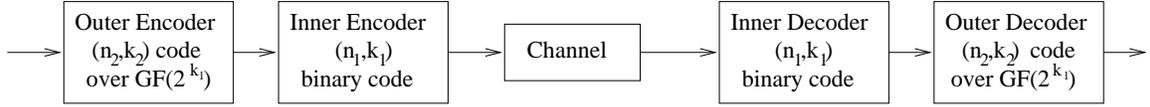

Fig. 5. A concatenated code.

The two-stage decoder shown in Figure 5 is not optimum, but is capable of correcting a wide variety of error patterns. For example, any error pattern that causes at most one error in each of the inner codewords will be corrected. In addition, if bursty errors cause one or two inner codewords to be decoded incorrectly, they will appear as correctable symbol errors to the outer decoder. The overall result is a long, powerful code with a simple, suboptimum decoder that can correct many combinations of burst and random errors. Forney showed that with a proper choice of the constituent codes, concatenated coding schemes could operate at any code rate up to the Shannon limit with exponentially decreasing error probability, but only polynomial decoding complexity.

Concatenation can also be applied to convolutional codes. In fact, the most common concatenated code used in practice is one developed in the 1970s as a NASA standard (mentioned in Section III-E). It consists of an inner rate-1/2, 64-state convolutional code with minimum distance[8] $d = 10$ along with an outer $(255, 223, 33)$ RS code over $\mathbb{F}_{256}$. The inner decoder uses soft-decision Viterbi decoding, while the outer decoder uses the hard-decision Berlekamp-Massey algorithm. Also, since the decoding errors made by the Viterbi algorithm tend to be bursty, a symbol interleaver is inserted between the two encoders, and a de-interleaver between the two decoders.

In the late 1980s, a more complex concatenated coding scheme with iterative decoding was proposed by Erik Paaske [66], and independently by Oliver Collins [67], to improve the performance of the NASA concatenated coding standard. Instead of a single outer Reed-Solomon (RS) code, Paaske and Collins proposed to use several outer RS codes of different rates. After one round of decoding, the outputs of the strongest (lowest-rate) RS decoders may be deemed to be reliable, and thus may be fed back to the inner (Viterbi) convolutional decoder as known bits for another round of decoding. Performance improvements of about 1.0 dB were achieved after a few iterations. This scheme was used to rescue the 1992 Galileo mission (see also Section IV-F). Also, in retrospect, its use of iterative decoding with a concatenated code may be seen as a precursor of turbo codes (see, for example, the paper by Hagenauer *et al.* [68]).

### E. Trellis decoding of block codes

A convolutional code may be viewed as the output sequence of a discrete-time, finite-state system. By rolling out the state-transition diagram of such a system in time, we get a picture called a *trellis diagram*, which explicitly displays every possible state sequence, and also every possible output sequence (if state transitions are labelled by the corresponding outputs). With such a trellis representation of a convolutional code, it becomes obvious that on a memoryless channel the Viterbi algorithm is a maximum-likelihood sequence detection algorithm [56].

The success of VA decoding of convolutional codes led to the idea of representing a block code by a (necessarily time-varying) trellis diagram with as few states as possible, and then using the VA to decode it. Another fundamental contribution of the BCJR paper [63] was to show that every $(n, k, d)$ binary linear code may be represented by a trellis diagram with at most $\min\{2^k, 2^{n-k}\}$ states.[9]

The subject of minimal trellis representations of block codes became an active research area during the 1990s. Given a linear block code with a fixed coordinate ordering, it turns out that there is a unique minimal trellis representation; however, finding the best coordinate ordering is an NP-hard problem. Nonetheless, optimal coordinate orderings for many classes of linear block codes have been found. In particular, the optimum coordinate ordering for Golay and Reed-Muller codes is known, and the resulting trellis diagrams are rather nice. On the other hand, the state complexity of any class of "good" block codes must increase exponentially as $n \to \infty$. An excellent summary of this field by Vardy appears in [70].

---

[8]The minimum distance between infinite code sequences in a convolutional code is also known as the *free distance*.

[9]This result is usually attributed to a subsequent paper by Wolf [69].

In practice, this approach was superseded by the advent of turbo and LDPC codes, to be discussed in Section VI.

### F. History of coding for deep-space applications

The deep-space communications application is the arena in which the most powerful coding schemes for the power-limited AWGN channel have been first deployed, because:

- The only noise is AWGN in the receiver front end;
- Bandwidth is effectively unlimited;
- Fractions of a dB have huge scientific and economic value;
- Receiver (decoding) complexity is effectively unlimited.

As we have already noted, for power-limited AWGN channels, there is a negligible penalty to using binary codes with binary modulation rather than more general modulation schemes.

The first coded scheme to be designed for space applications was a simple $(32, 6, 16)$ biorthogonal code for the Mariner missions (1969), which can be optimally soft-decision decoded using a fast Hadamard transform. Such a scheme can achieve a nominal coding gain of 3 (4.8 dB). At a target bit error probability of $P_b(E) \approx 5 \cdot 10^{-3}$, the real coding gain achieved was only about 2.2 dB.

The first coded scheme actually to be launched into space was a rate-1/2 convolutional code with constraint length[10] $\nu = 20$ ($2^{20}$ states) for the Pioneer 1968 mission [3]. The receiver used 3-bit-quantized soft decisions and sequential decoding implemented on a general-purpose 16-bit minicomputer with a 1 MHz clock rate. At a rate of 512 b/s, the real coding gain achieved at $P_b(E) \approx 5 \cdot 10^{-3}$ was about 3.3 dB.

During the 1970s, as noted in Sections III-E and IV-D, the NASA standard became a concatenated coding scheme based on a rate-1/2, 64-state inner convolutional code and a $(255, 223, 33)$ Reed-Solomon outer code over $\mathbb{F}_{256}$. The overall rate of this code is $0.437$, and it achieves an impressive 7.3 dB real coding gain at $P_b(E) \approx 10^{-5}$; *i.e.*, its gap to capacity ($\text{SNR}_{\text{norm}}$) is only about 2.5 dB (see Figure 6).

When the primary antenna failed to deploy on the Galileo mission (*circa* 1992), an elaborate concatenated coding scheme using a rate-1/6, $2^{14}$-state inner convolutional code with a Big Viterbi Decoder (BVD) and a set of variable-strength RS outer codes was reprogrammed into the spacecraft computers (see Section IV-D). This scheme was able to achieve $P_b(E) \approx 2 \cdot 10^{-7}$ at $E_b/N_0 \approx 0.8$ dB, for a real coding gain of about 10.2 dB.

Finally, within the last decade, turbo codes and LDPC codes for deep-space communications have been developed to get within 1 dB of the Shannon limit, and these are now becoming industry standards (see Section VI-H).

For a more comprehensive history of coding for deep-space channels, see [71].

### V. Codes for bandwidth-limited channels

Most work on channel coding has focussed on binary codes. However, on a bandwidth-limited AWGN channel, in order to obtain a spectral efficiency $\eta > 2$ b/s/Hz, some kind of nonbinary coding must be used.

Early work, primarily theoretical, focussed on lattice codes, which in many respects are analogous to binary linear block codes. The practical breakthrough in this field came with Ungerboeck's invention of trellis-coded modulation, which is similarly analogous to convolutional coding.

---

[10]The constraint length $\nu$ is the dimension of the state space of a convolutional encoder; the number of states is thus $2^\nu$.

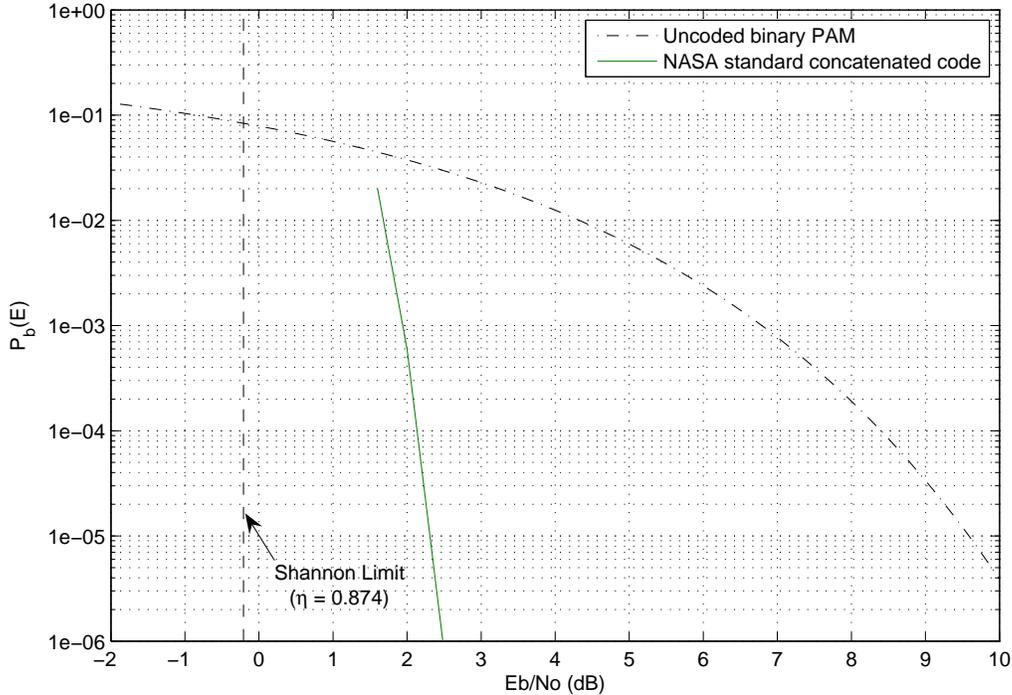

Fig. 6. $P_b(E)$ vs. $E_b/N_0$ for the NASA standard concatenated code, compared to uncoded PAM and the Shannon limit for $\eta = 0.874$.

### A. Coding for the bandwidth-limited AWGN channel

Coding schemes for a bandwidth-limited AWGN channel typically use two-dimensional quadrature amplitude modulation (QAM). A sequence of QAM symbols may be sent through a channel of bandwidth $W$ at a symbol rate up to the Nyquist limit of $W$ QAM symbols per second. If the information rate is $\eta$ bits per QAM symbol, then the nominal spectral efficiency is also $\eta$ b/s/Hz.

An uncoded baseline scheme is simply to use a square $M \times M$ QAM constellation, where $M$ is even, typically a power of two. The information rate is then $\eta = \log_2 M^2$ bits per QAM symbol. The average energy of such a constellation is easily shown to be

$$E_s = \frac{(M^2 - 1)d^2}{6} = \frac{(2^\eta - 1)d^2}{6},$$

where $d$ is the minimum Euclidean distance between constellation points. Since $\mathrm{SNR} = E_s/N_0$, it is then straightforward to show that with optimum modulation and detection the probability of error per QAM symbol is

$$P_s(E) \approx 4\,Q(\sqrt{3 \cdot \mathrm{SNR}_{\mathrm{norm}}}),$$

where $Q(x)$ is again the Gaussian probability of error function.

This baseline performance curve of $P_s(E)$ vs. $\mathrm{SNR}_{\mathrm{norm}}$ for uncoded QAM transmission is plotted in Figure 7. For example, in order to achieve a symbol error probability of $P_s(E) \approx 10^{-5}$, we must have $\mathrm{SNR}_{\mathrm{norm}} \approx 7$ (8.5 dB) for uncoded QAM transmission.

We recall from Section II that the Shannon limit on $\mathrm{SNR}_{\mathrm{norm}}$ is 1 (0 dB), so the gap to capacity is about 8.5 dB at $P_s(E) \approx 10^{-5}$. Thus the maximum possible coding gain is somewhat smaller in the bandwidth-limited regime than in the power-limited regime. Furthermore, as we will discuss next, in the bandwidth-limited regime the Shannon limit on $\mathrm{SNR}_{\mathrm{norm}}$ with no shaping is $\pi e/6$ (1.53 dB), so the maximum possible coding gain with no shaping at $P_s(E) \approx 10^{-5}$ is only about 7 dB. These two limits are also shown on Figure 7.

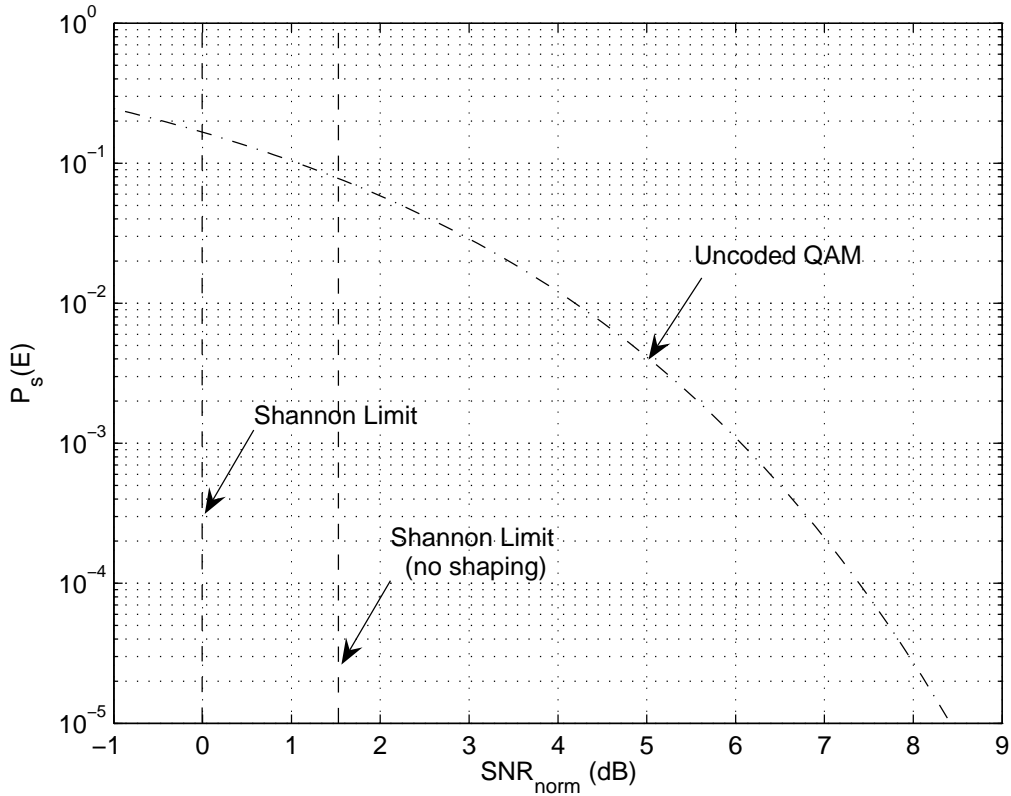

Fig. 7. $P_s(E)$ *vs.* $\mathrm{SNR}_{\mathrm{norm}}$ for uncoded QAM, compared to Shannon limits on $\mathrm{SNR}_{\mathrm{norm}}$ with and without shaping.

We now briefly discuss shaping. The set of all $n$-tuples of constellation points from a square QAM constellation is the set of all points on a $d$-spaced rectangular grid that lie within a $2n$-cube in real $2n$-space $\mathbb{R}^{2n}$. The average energy of this $2n$-dimensional constellation could be reduced if instead the constellation consisted of all points on the same grid that lie within a $2n$-sphere of the same volume, which would comprise approximately the same number of points. The reduction in average energy of a $2n$-sphere relative to a $2n$-cube of the same volume is called the *shaping gain* $\gamma_s(S_{2n})$ of a $2n$-sphere. As $n \to \infty$, $\gamma_s(S_n) \to \pi e/6$ (1.53 dB).

For large signal constellations, shaping can be implemented more or less independently of coding, and shaping gain is more or less independent of coding gain. The Shannon limit essentially assumes $n$-sphere shaping with $n \to \infty$, and therefore incorporates 1.53 dB of shaping gain (over an uncoded square QAM constellation). In the bandwidth-limited regime, coding without shaping can therefore get only to within 1.53 dB of the Shannon limit; the remaining 1.53 dB can be obtained by shaping and only by shaping.

We do not have space to discuss shaping schemes in this paper. It turns out that obtaining shaping gains on the order of 1 dB is not very hard, so nowadays most practical schemes for the bandwidth-limited Gaussian channel incorporate shaping. For example, the V.34 modem (see Section V-D) incorporates a 16-dimensional "shell mapping" shaping scheme whose shaping gain is about 0.8 dB.

The performance curve of any practical coding scheme that improves on uncoded QAM must lie between the relevant Shannon limit and the uncoded QAM curve. Thus Figure 7 defines the "playing field" for coding and shaping in the bandwidth-limited regime. The *real coding gain* of a coding scheme at a given symbol error probability $P_s(E)$ will be defined as the difference (in dB) between the $\mathrm{SNR}_{\mathrm{norm}}$ required to obtain that $P_s(E)$ with coding, but no shaping, *vs.* without coding (uncoded QAM). Thus the maximum possible real coding gain at $P_s(E) \approx 10^{-5}$ is about 7 dB.

Again, for moderate-complexity coding, it can often be assumed that the error probability is dominated by the probability of making an error to one of the nearest-neighbor codewords. Under this assumption, using a union bound estimate [75], [76], it is easily shown that with optimum decoding, the probability of decoding error per QAM symbol is well approximated by

$$P_s(E) \approx (2N_d/n) \; Q\left(\sqrt{3d^2 2^{-\rho}\mathrm{SNR}_{\mathrm{norm}}}\right) = (2N_d/n) \; Q\left(\sqrt{3\gamma_c\mathrm{SNR}_{\mathrm{norm}}}\right),$$

where $d^2$ is the minimum squared Euclidean distance between code sequences (assuming an underlying QAM constellation with minimum distance 1 between signal points), $2N_d/n$ is the number of code sequences at the minimum distance per QAM symbol, and $\rho$ is the redundancy of the coding scheme (the difference between the actual and maximum possible data rates with the underlying QAM constellation) in bits per two dimensions. The quantity $\gamma_c = d^2 2^{-\rho}$ is called the *nominal coding gain* of the coding scheme. The real coding gain is usually slightly less than the nominal coding gain, due to the effect of the "error coefficient" $2N_d/n$.

### B. Spherical lattice codes

It is clear from the proof of Shannon's capacity theorem for the AWGN channel that an optimal code for a bandwidth-limited AWGN channel consists of a dense packing of signal points within an $n$-sphere in a high-dimensional Euclidean space $\mathbb{R}^n$.

Finding the densest packings in $\mathbb{R}^n$ is a longstanding mathematical problem. Most of the densest known packings are lattices [72]– *i.e.,* packings that have a group property. Notable lattice packings include the integer lattice $\mathbb{Z}$ in one dimension, the hexagonal lattice $A_2$ in two dimensions, the Gosset lattice $E_8$ in eight dimensions, and the Leech lattice $\Lambda_{24}$ in 24 dimensions.

Therefore, from the very earliest days, there have been proposals to use spherical lattice codes as codes for the bandwidth-limited AWGN channel, notably by de Buda [73] and Lang in Canada. Lang proposed an $E_8$ lattice code for telephone-line modems to a CCITT international standards committee in the mid-70s, and actually built a Leech lattice modem in the late 1980s [74].

By the union bound estimate, the probability of error per two-dimensional symbol of a spherical lattice code based on an $n$-dimensional lattice $\Lambda$ on an AWGN channel with minimum-distance decoding may be estimated as

$$P_s(E) \approx 2K_{\min}(\Lambda)/n \; Q\left(\sqrt{3\gamma_c(\Lambda)\gamma_s(S_n)\mathrm{SNR}_{\mathrm{norm}}}\right),$$

where $K_{\min}(\Lambda)$ is the kissing number (number of nearest neighbors) of the lattice $\Lambda$, $\gamma_c(\Lambda)$ is the nominal coding gain (Hermite parameter) of $\Lambda$, and $\gamma_s(S_n)$ is the shaping gain of an $n$-sphere. Since $P_s(E) \approx 4Q\left(\sqrt{3\mathrm{SNR}_{\mathrm{norm}}}\right)$ for a square two-dimensional QAM constellation, the real coding gain of a spherical lattice code over a square QAM constellation is the combination of the nominal coding gain $\gamma_c(\Lambda)$ and the shaping gain $\gamma_s(S_n)$, minus a $P_s(E)$-dependent factor due to the larger "error coefficient" $2K_{\min}(\Lambda)/n$.

For example (see [76]), the Gosset lattice $E_8$ has a nominal coding gain of 2 (3 dB); however, $K_{\min}(E_8) = 240$, so with no shaping

$$P_s(E) \approx 60 \; Q\left(\sqrt{6\mathrm{SNR}_{\mathrm{norm}}}\right),$$

which is plotted in Figure 8. We see that the real coding gain of $E_8$ is only about 2.2 dB at $P_s(E) \approx 10^{-5}$. The Leech lattice $\Lambda_{24}$ has a nominal coding gain of 4 (6 dB); however, $K_{\min}(\Lambda_{24}) = 196560$, so with no shaping

$$P_s(E) \approx 16380 \; Q\left(\sqrt{12\mathrm{SNR}_{\mathrm{norm}}}\right),$$

also plotted in Figure 8. We see that the real coding gain of $\Lambda_{24}$ is only about 3.6 dB at $P_s(E) \approx 10^{-5}$. Spherical shaping in 8 or 24 dimensions would contribute a shaping gain of about 0.75 dB or 1.1 dB, respectively.

For a detailed discussion of lattices and lattice codes, see the book by Conway and Sloane [72].

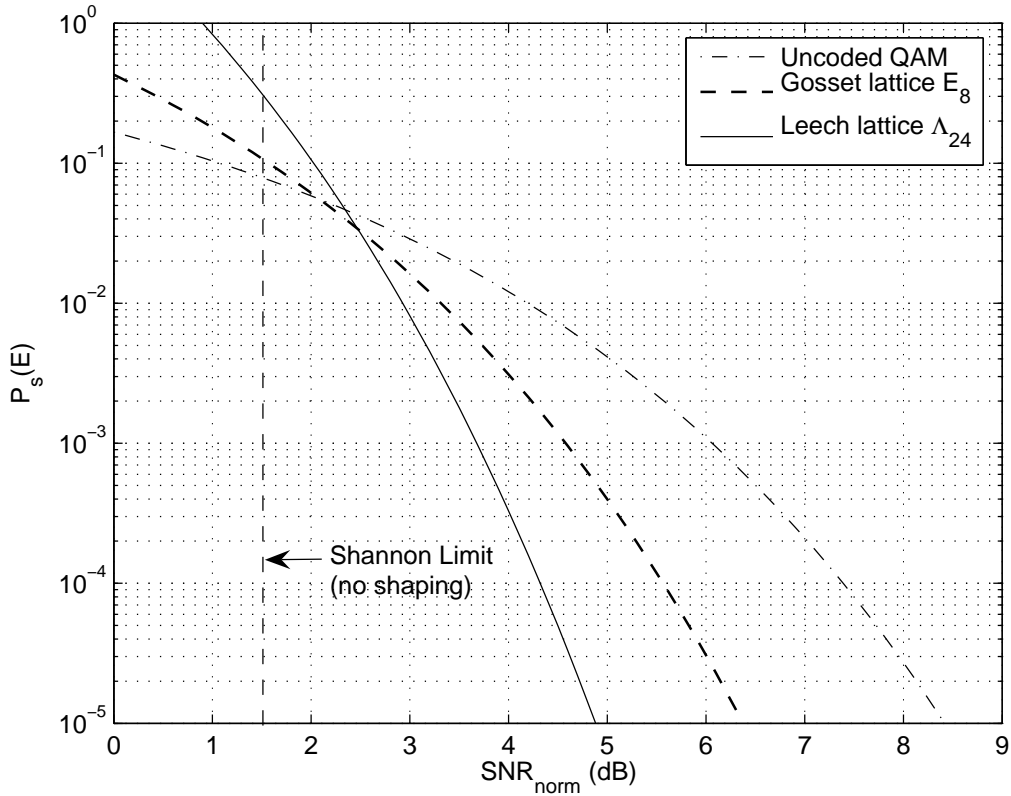

Fig. 8. $P_s(E)$ *vs.* $\mathrm{SNR_{norm}}$ for Gosset lattice $E_8$ and Leech lattice $\Lambda_{24}$ with no shaping, compared to uncoded QAM and the Shannon limit on $\mathrm{SNR_{norm}}$ without shaping.

### C. Trellis-coded modulation

The big breakthrough in practical coding for bandwidth-limited channels was Gottfried Ungerboeck's invention of trellis-coded modulation (TCM), originally conceived in the 1970s, but not published until 1982 [77].

Ungerboeck realized that in the bandwidth-limited regime, the redundancy needed for coding should be obtained by expanding the signal constellation while keeping the bandwidth fixed, rather than by increasing the bandwidth while keeping a fixed signal constellation, as is done in the power-limited regime. From capacity calculations, he showed that doubling the signal constellation should suffice— *e.g.,* using a 32-QAM rather than a 16-QAM constellation. Ungerboeck invented clever trellis codes for such expanded constellations, using minimum Euclidean distance rather than Hamming distance as the design criterion.

As with convolutional codes, trellis codes may be optimally decoded by a VA decoder, whose decoding complexity is proportional to the number of states in the encoder.

Ungerboeck showed that effective coding gains of 3 to 4 dB could be obtained with simple 4- to 8-state trellis codes, with no bandwidth expansion. An 8-state two-dimensional (2D) QAM trellis code due to Lee-Fang Wei [79] (with a nonlinear twist to make it "rotationally invariant") was soon incorporated into the V.32 voice-grade telephone-line modem standard (see Section V-D). The nominal (and real) coding gain of this 8-state 2D code is $\gamma_c = 5/2 = 2.5$ (3.97 dB); its performance curve is approximately $P_s(E) \approx 4\,Q(\sqrt{7.5\,\mathrm{SNR_{norm}}})$, plotted in Figure 9. Later standards such as V.34 have used a 16-state 4D trellis code of Wei [80] (see Section V-D), which has less redundancy ($\rho = \frac{1}{2}$ *vs.* $\rho = 1$), a nominal coding gain of $\gamma_c = 4/\sqrt{2} = 2.82$ (4.52 dB), and performance $P_s(E) \approx 12\,Q(\sqrt{8.49\,\mathrm{SNR_{norm}}})$, also plotted in Figure 9. We see that its real coding gain at $P_s(E) \approx 10^{-5}$ is about 4.2 dB.

Trellis codes have proved to be more attractive than lattice codes in terms of performance *vs.* complexity, just as convolutional codes have been preferred to block codes. Nonetheless, the signal constellations used for trellis codes have generally been based on simple lattices, and their "subset partitioning" is often best understood as being based on a sublattice chain. For example, the V.32 code uses a QAM constellation based on the two-dimensional integer lattice $\mathbb{Z}^2$, with an 8-way partition based on the sublattice chain $\mathbb{Z}^2/R_2\mathbb{Z}^2/2\mathbb{Z}^2/2R_2\mathbb{Z}^2$, where $R_2$ is a scaled rotation operator. The Wei 4D 16-state code uses a constellation based on the 4-dimensional integer lattice $\mathbb{Z}^4$, with an 8-way partition based on the sublattice chain $\mathbb{Z}^4/D_4/R_4\mathbb{Z}^4/R_4D_4$, where $D_4$ is the 4-dimensional "checkerboard lattice," and $R_4$ is a 4D extension of $R_2$.

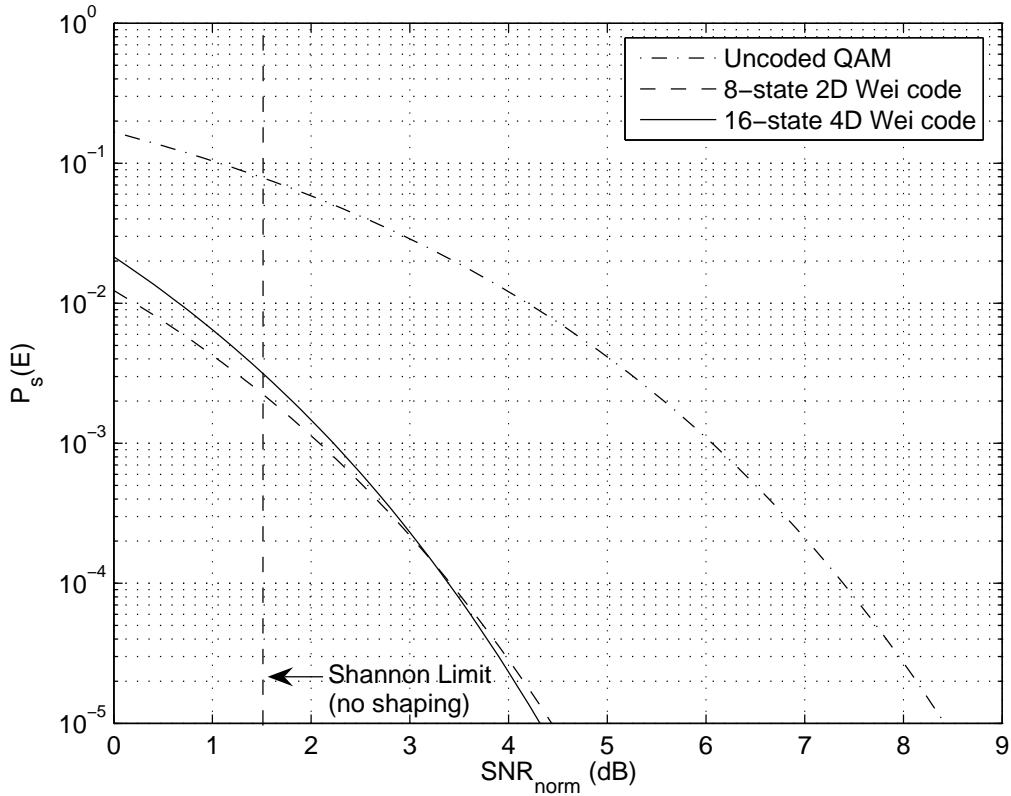

Fig. 9. $P_s(E)$ *vs.* $\mathrm{SNR_{norm}}$ for 8-state 2D and 16-state 4D Wei trellis codes with no shaping, compared to uncoded QAM and the Shannon limit on $\mathrm{SNR_{norm}}$ without shaping.

In 1977, Imai and Hirakawa introduced a related concept, called multilevel coding [78]. In this approach, an independent binary code is used at each stage of a chain of 2-way partitions, such as $\mathbb{Z}^2/R_2\mathbb{Z}^2/2\mathbb{Z}^2/2R_2\mathbb{Z}^2$. By information-theoretic arguments, it can be shown that multilevel coding suffices to approach the Shannon limit [124]. However, TCM has been the preferred approach in practice.

### D. History of coding for modem applications

For several decades, the telephone channel was the arena in which the most powerful coding and modulation schemes for the bandwidth-limited AWGN channel were first developed and deployed, because:

- At that time, the telephone channel was fairly well modeled as a bandwidth-limited AWGN channel;
- One dB had significant commercial value;
- Data rates were low enough that a considerable amount of processing could be done per bit.

The first international standard to use coding was the V.32 standard (1986) for 9600 b/s transmission over the public switched telephone network (PSTN) (later raised to 14.4 kb/s in V.32*bis*). This modem used an 8-state, 2D rotationally invariant Wei trellis code to achieve a real coding gain of about 3.5 dB with a 32-QAM (later 128-QAM in V.32*bis*) constellation at 2400 symbols/s— *i.e.,* a nominal bandwidth of 2400 Hz.

The "ultimate modem standard" was V.34 (1994) for transmission at up to 28.8 kb/s over the PSTN (later raised to 33.6 kb/s in V.34*bis*). This modem used a 16-state, 4D rotationally invariant Wei trellis code to achieve a coding gain of about 4.0 dB with a variable-sized QAM constellation with up to 1664 points. An optional 32-state, 4D trellis code with an additional coding gain of 0.3 dB and four times (4x) the decoding complexity and a 64-state, 4D code with a further 0.15 dB coding gain and a further 4x increase in complexity were also specified. A 16D "shell mapping" constellation shaping scheme provided an additional gain of about 0.8 dB. A variable symbol rate of up to 3429 symbols/s was used, with symbol rate and data rate selection determined by "line probing" of individual channels.

However, the V.34 standard was shortly superseded by V.90 (1998) and V.92 (2000), which allow users to send data directly over the 56 or 64 kb/s digital backbones that are now nearly universal in the PSTN. Neither V.90 nor V.92 uses coding, because of the difficulty of achieving coding gain on a digital channel.

Currently, coding techniques similar to those of V.34 are used in higher-speed wireline modems, such as digital subscriber line (DSL) modems, as well as on digital cellular wireless channels. Capacity-approaching coding schemes are now normally included in new wireless standards. In other words, bandwidth-limited coding has moved to these newer, higher-bandwidth settings.

## VI. THE TURBO REVOLUTION

In 1993, at the IEEE International Conference on Communications (ICC) in Geneva, Berrou, Glavieux, and Thitimajshima [81] stunned the coding research community by introducing a new class of "turbo codes" that purportedly could achieve near-Shannon-limit performance with modest decoding complexity. Comments to the effect of "It can't be true; they must have made a 3 dB error" were widespread.[11] However, within the next year various laboratories confirmed these astonishing results, and the "turbo revolution" was launched.

Shortly thereafter, codes similar to Gallager's LDPC codes were discovered independently by MacKay at Cambridge [82], [83] and by Spielman at MIT [84], [85], along with low-complexity iterative decoding algorithms. MacKay showed that in practice moderate-length LDPC codes ($10^3$-$10^4$ bits) could attain near-Shannon-limit performance, whereas Spielman showed that in theory, as $n \to \infty$, they could approach the Shannon limit with linear decoding complexity. These results kicked off a similar explosion of research on LDPC codes, which are currently seen as competitors to turbo codes in practice.

In 1995, Wiberg showed in his doctoral thesis at Linköping [86], [87] that both of these classes of codes could be understood as instances of "codes on sparse graphs," and that their decoding algorithms could be understood as instances of a general iterative APP decoding algorithm called the "sum-product algorithm." Late in his thesis work, Wiberg discovered that many of his results had previously been found by Tanner [88], in a largely forgotten 1981 paper. Wiberg's rediscovery of Tanner's work opened up a new field, called "codes on graphs."

In this section we will discuss the various historical threads leading to and springing from these watershed events of the mid-90's, which have proved effectively to answer the challenge laid down by Shannon in 1948.

### A. Precursors

As we have discussed in previous sections, certain elements of the turbo revolution had been appreciated for a long time. It had been known since the early work of Elias that linear codes were as good as general codes. Information theorists also understood that maximizing the minimum distance was not the key to getting to capacity; rather, codes should be "random-like," in the sense that the distribution of distances from a typical codeword to all other codewords should resemble the distance distribution in a random code. These principles were already evident in Gallager's monograph on LDPC codes [49]. Gérard Battail, whose work inspired Berrou and Glavieux, was a long-time advocate of seeking "random-like" codes (see, *e.g.,* [89]).

---

[11]Although both were professors, neither Berrou nor Glavieux had completed a doctoral degree.

Another element of the turbo revolution whose roots go far back is the use of soft decisions (reliability information) not only as input to a decoder, but also in the internal workings of an iterative decoder. Indeed, by 1962 Gallager had already developed the modern APP decoder for decoding LDPC codes and had shown that retaining soft-decision (APP) information in iterative decoding was useful even on a hard-decision channel such as a BSC.

The idea of using soft-input, soft-output (SISO) decoding in a concatenated coding scheme originated in papers by Battail [90] and by Joachim Hagenauer and Peter Hoeher [91]. They proposed a SISO version of the Viterbi algorithm, called the soft-output Viterbi algorithm (SOVA). In collaboration with John Lodge, Hoeher and Hagenauer extended their ideas to iterating separate SISO APP decoders [92]. Moreover, at the same 1993 ICC at which Berrou *et al.* introduced turbo codes and first used the term "extrinsic information" (see discussion in next section), a paper by Lodge *et al.* [93] also included the idea of "extrinsic information". By this time the benefits of retaining soft information throughout the decoding process had been clearly appreciated; see, for example, Battail [90] and Hagenauer [94]. We have already noted in Section IV-D that similar ideas had been developed at about the same time in the context of NASA's iterative decoding scheme for concatenated codes.

### B. The turbo code breakthrough

The invention of turbo codes began with Alain Glavieux's suggestion to his colleague Claude Berrou, a professor of VLSI circuit design, that it would be interesting to implement the SOVA decoder in silicon. While studying the principles underlying the SOVA decoder, Berrou was struck by Hagenauer's statement that "a SISO decoder is a kind of SNR amplifier." As a physicist, Berrou wondered whether the SNR could be further improved by repeated decoding, using some sort of "turbo-type" iterative feedback. As they say, the rest is history.

The original turbo encoder design introduced in [81] is shown in Figure 10. An information sequence $\mathbf{u}$ is encoded by an ordinary rate-1/2, 16-state, systematic recursive convolutional encoder to generate a first parity bit sequence; the same information bit sequence is then scrambled by a large pseudorandom interleaver $\pi$ and encoded by a second, identical rate-1/2 systematic convolutional encoder to generate a second parity bit sequence. The encoder transmits all three sequences, so the overall encoder has rate 1/3. (This is now called the "parallel concatenation" of two codes, in contrast with the original kind of concatenation, now called "serial.")

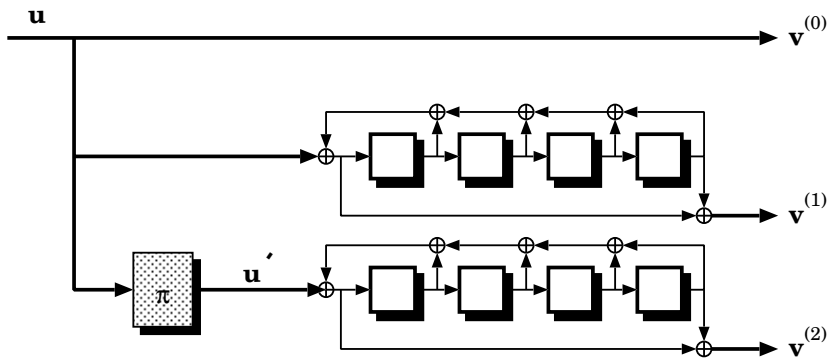

Fig. 10. A parallel concatenated rate-1/3 turbo encoder.

The use of recursive (feedback) convolutional encoders and an interleaver turn out to be critical for making a turbo code somewhat "random-like." If a nonrecursive encoder were used, then a single nonzero information bit would necessarily generate a low-weight code sequence. It was soon shown by Benedetto *et al.* [95] and by Perez *et al.* [96] that the use of a length-$N$ interleaver effectively reduces the number of low-weight codewords by a factor of $N$. However, turbo codes nevertheless have relatively poor minimum distance. Indeed, Breiling has shown that the minimum distance of turbo codes grows only logarithmically with the interleaver length $N$ [97].

The iterative turbo decoding system is shown in Figure 11. Decoders 1 and 2 are APP (BCJR) decoders for the two constituent convolutional codes, $\Pi$ is the same permutation as in the encoder, and $\Pi^{-1}$ is the inverse permutation. Berrou *et al.* discovered that the key to achieving good iterative decoding performance is the removal of the "intrinsic information" from the output APPs $L^{(i)}(u_l)$, resulting in "extrinsic" APPs $L_e^{(i)}(u_l)$, which are then passed as *a priori* inputs to the other decoder. "Intrinsic information" represents the soft channel outputs $L_c r_l^{(i)}$ and the *a priori* inputs already known prior to decoding, while "extrinsic information" represents additional knowledge learned about an information bit during an iteration. The removal of "intrinsic information" has the effect of reducing correlations from one decoding iteration to the next, thus allowing improved performance with an increasing number of iterations.[12] (See Chapter 16 of [44] for more details.) The iterative feedback of the "extrinsic" APPs recalls the feedback of exhaust gases in a turbo-charged engine.

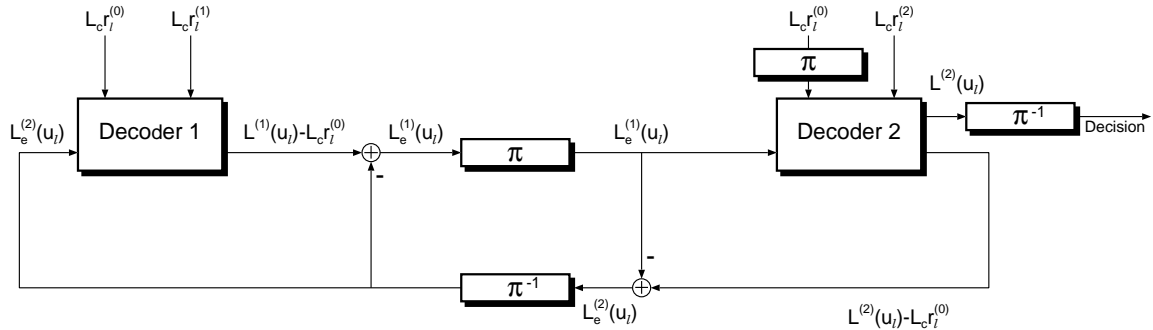

Fig. 11.   Iterative decoder for a parallel concatenated turbo code.

The performance on an AWGN channel of the turbo code and decoder of Figures 10 and 11, with an interleaver length of $N = 2^{16}$, after "puncturing" (deleting symbols) to raise the code rate to 1/2 ($\eta = 1$ b/s/Hz), is shown in Figure 12. At $P_b(E) \approx 10^{-5}$, performance is about 0.7 dB from the Shannon limit for $\eta = 1$, or only 0.5 dB from the Shannon limit for binary codes at $\eta = 1$. In contrast, the real coding gain of the NASA standard concatenated code is about 1.6 dB less, even though its rate is lower and its decoding complexity is about the same.

With randomly constructed interleavers, at values of $P_b(E)$ somewhat below $10^{-5}$, the performance curve of turbo codes typically flattens out, resulting in what has become known as an "error floor" (as seen in Figure 12, for example.) This happens because turbo codes do not have large minimum distances, so ultimately performance is limited by the probability of confusing the transmitted codeword with a near neighbor. Several approaches have been suggested to mitigate the error-floor effect. These include using serial concatenation rather than parallel concatenation (see, for example, [98] or [99]), the design of **structured** interleavers to improve the minimum distance (see, for example, [100] or [101]), or the use of multiple interleavers to eliminate low-weight codewords (see, for example, [102] or [103]). However, the fact that the minimum distance of turbo codes cannot grow linearly with block length implies that the ensemble of turbo codes is not asymptotically good, and that "error floors" cannot be totally avoided.

## C. Rediscovery of LDPC codes

Gallager's invention of LDPC codes and the iterative APP decoding algorithm was long before its time ("a bit of 21st-century coding that happened to fall in the 20th century"). His work was largely forgotten for more than 30 years. It is easy to understand why there was little interest in LDPC codes in the 60s and 70s, because these codes were much too complex for the technology of that time. It is not so easy to explain why they continued to be ignored by the coding community up to the mid-90s.

Shortly after the turbo code breakthrough, several researchers with backgrounds in computer science and physics rather than in coding rediscovered the power and efficiency of LDPC codes. In his thesis, Dan Spielman [84], [85] used LDPC codes based on expander graphs to devise codes with linear-time encoding and decoding algorithms and with respectable error performance. At about the same time and independently, David MacKay [83], [104] showed empirically that near-Shannon-limit performance could be obtained with long LDPC-type codes and iterative decoding.

---

[12]We note, however, that correlations do build up with iterations and that a saturation effect is eventually observed, where no further improvement is possible.

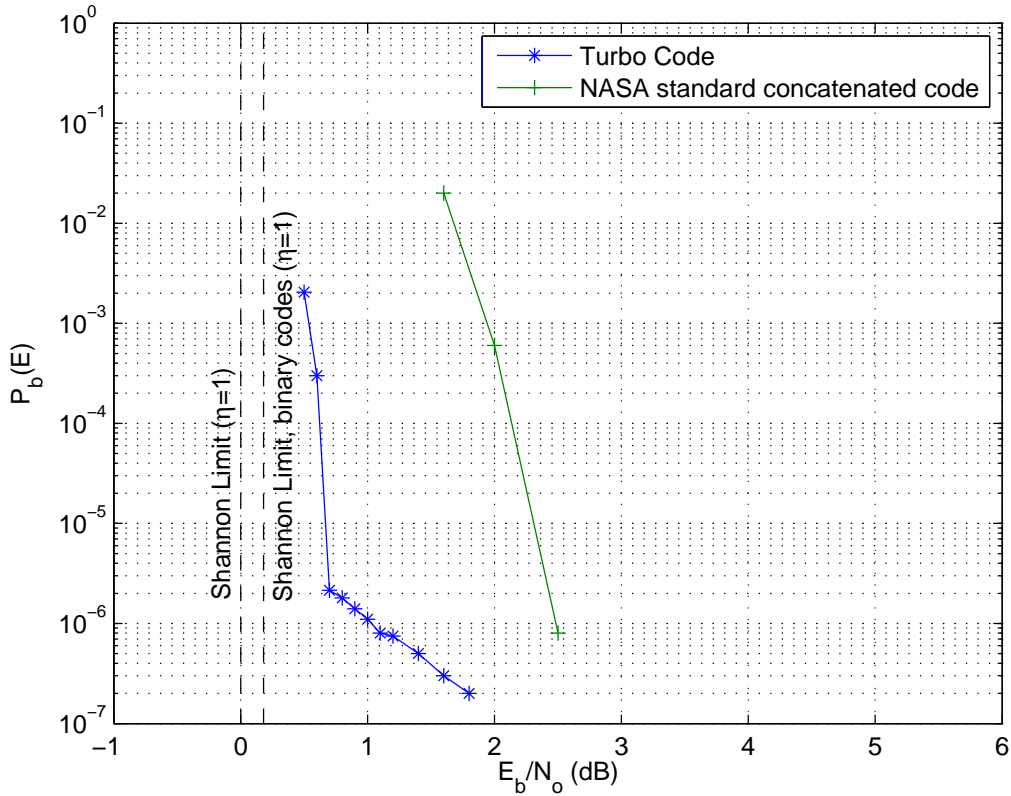

Fig. 12. Performance of a rate-1/2 turbo code with interleaver length $N = 2^{16}$, compared to the NASA standard concatenated code and the relevant Shannon limits for $\eta = 1$.

Given that turbo codes were already a hot topic, the rediscovery of LDPC codes kindled an explosion of interest in this field that has continued to this day.

An LDPC code is commonly represented by a bipartite graph as in Figure 13, introduced by Michael Tanner in 1981 [88], and now called a "Tanner graph." Each code symbol $y_k$ is represented by a node of one type, and each parity check by a node of a second type. A symbol node and a check node are connected by an edge if the corresponding symbol is involved in the corresponding check. In an LDPC code, the edges are sparse, in the sense that their number increases linearly with the block length $n$, rather than as $n^2$.

The impressive complexity results of Spielman were quickly applied by Alon and Luby [105] to the Internet problem of reconstructing large files in the presence of packet erasures. This work exploits the fact that on an erasure channel,[13] decoding linear codes is essentially a matter of solving linear equations, and becomes very efficient if it can be reduced to solving a series of equations, each of which involves a single unknown variable.

An important general discovery that arose from this work was the superiority of irregular LDPC codes. In a regular LDPC code, such as the one shown in Figure 13, all symbol nodes have the same degree (number of incident edges), and so do all check nodes. Luby *et al.* [106], [107] found that by using irregular graphs and optimizing the degree sequences (numbers of symbol and check nodes of each degree), they could approach the capacity of the erasure channel— *i.e.,* achieve small error probabilities at code rates of nearly 1 - $p$, where $p$ is the erasure probability. For example, a rate-1/2 LDPC code capable of correcting up to a fraction $p = 0.4955$ of erasures is described in [108]. "Tornado codes" of this type were commercialized by Digital Fountain, Inc. [109].

---

[13]On an erasure channel, transmitted symbols or packets are either received correctly or not at all; *i.e.,* there are no "channel errors."

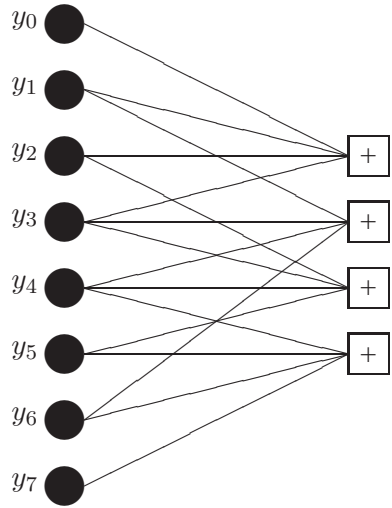

Fig. 13.   Tanner graph of the $(8, 4, 4)$ extended Hamming code.

More recently, it has been shown [110] that on any erasure channel, binary or nonbinary, it is possible to design LDPC codes that can approach capacity arbitrarily closely, in the limit as $n \to \infty$. The erasure channel is the only channel for which such a result has been proved.

Building on the analytical techniques developed for Tornado codes, Richardson, Urbanke *et al.* [111], [112] used a technique called "density evolution" to design long irregular LDPC codes that for all practical purposes achieve the Shannon limit on binary AWGN channels.

Given an irregular binary LDPC code with arbitrary degree sequences, they showed that the evolution of probability densities on a binary-input memoryless symmetric (BMS) channel using an iterative sum-product (or similar) decoder can be analyzed precisely. They proved that error-free performance could be achieved below a certain threshold, for very long codes and large numbers of iterations. Degree sequences may then be chosen to optimize the threshold. By simulations, they showed that codes designed in this way could clearly outperform turbo codes for block lengths of the order of $10^5$ or more.

Using this approach, Chung *et al.* [113] designed several rate-1/2 codes for the AWGN channel, including one whose theoretical threshold approached the Shannon limit within 0.0045 dB, and another whose simulated performance with a block length of $10^7$ approached the Shannon limit within 0.040 dB at an error rate of $P_b(E) \approx 10^{-6}$, as shown in Figure 14. It is rather surprising that this close approach to the Shannon limit required no extension of Gallager's LDPC codes beyond irregularity. The former (threshold-optimized) code had symbol node degrees $\{2, 3, 6, 7, 15, 20, 50, 70, 100, 150, 400, 900, 2000, 3000, 6000, 8000\}$, with average degree $\overline{d}_\lambda = 9.25$, and check node degrees $\{18, 19\}$, with average degree $\overline{d}_\rho = 18.5$. The latter (simulated) code had symbol degrees $\{2, 3, 6, 7, 18, 19, 55, 56, 200\}$, with $\overline{d}_\lambda = 6$, and all check degrees equal to 12.

In current research, more structured LDPC codes are being sought for shorter block lengths, of the order of 1000. The original work of Tanner [88] included several algebraic constructions of codes on graphs. Algebraic structure may be preferable to a pseudo-random structure for implementation and may allow control over important code parameters such as minimum distance, as well as graph-theoretic variables such as expansion and girth.[14] The most impressive results are perhaps those of [114], in which it is shown that certain classical finite-geometry codes and their extensions can produce good LDPC codes. High-rate codes with lengths up to 524,256 have been constructed and shown to perform within 0.3 dB of the Shannon limit.

---

[14]Expansion and girth are properties of a graph that relate to its suitability for iterative decoding.

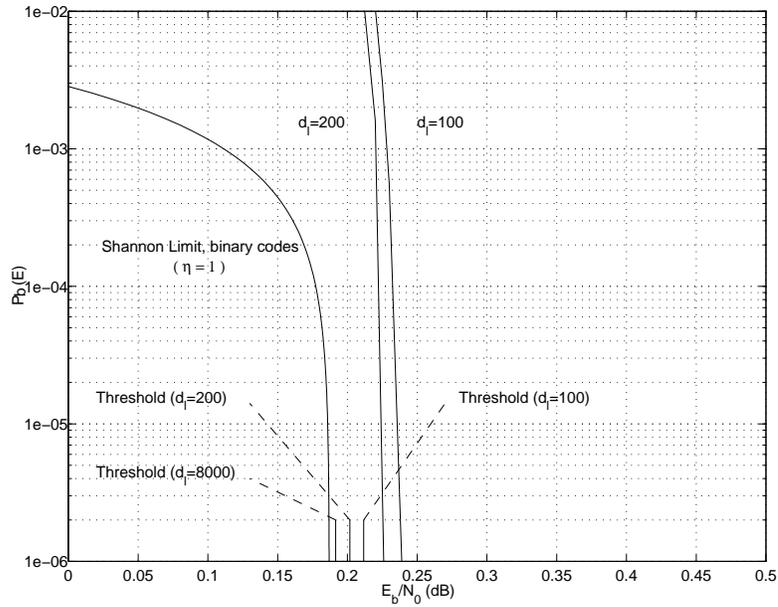

Fig. 14. Performance of optimized rate-$\frac{1}{2}$ irregular LDPC codes: asymptotic analysis with maximum symbol degree $d_l = 100, 200, 8000$, and simulations with maximum symbol degree $d_l = 100, 200$ and $n = 10^7$ [113].

### D. RA codes and other variants

Divsalar, McEliece *et al.* [115] proposed "repeat-accumulate" (RA) codes in 1998 as simple "turbo-like" codes for which one could prove coding theorems. An RA code is generated by the serial concatenation of a simple $(n, 1, n)$ repetition code, a large pseudo-random interleaver $\Pi$, and a simple 2-state rate-1/1 convolutional "accumulator" code with input-output equation $y_k = x_k + y_{k-1}$, as shown in Figure 15.

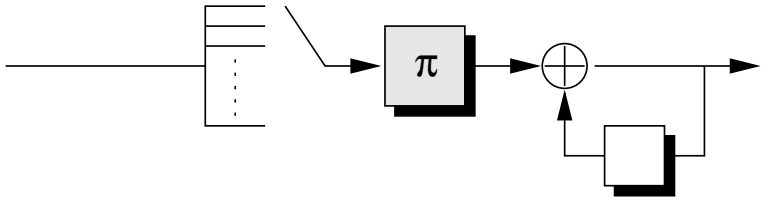

Fig. 15. A repeat-accumulate (RA) encoder.

The performance of RA codes turned out to be remarkably good, within about 1.5 dB of the Shannon limit— *i.e.,* better than that of the best coding schemes known prior to turbo codes.

Other authors have proposed equally simple codes with similar or even better performance. For example, RA codes have been extended to "accumulate-repeat-accumulate" (ARA) codes [116], which have even better performance. Ping and Wu [117] proposed "concatenated tree codes" comprising $M$ two-state trellis codes interconnected by interleavers, which exhibit performance almost identical to turbo codes of equal block length, but with an order of magnitude less complexity (see also Massey and Costello [118]). It seems that there are many ways that simple codes can be interconnected by large pseudo-random interleavers and decoded with the sum-product algorithm so as to yield near-Shannon-limit performance.

### E. Fountain (rateless) codes

*Fountain codes*, or "rateless codes," are a new class of codes designed for channels without feedback whose statistics are not known *a priori*; *e.g.,* Internet packet channels where the probability $p$ of packet erasure is unknown.

The "fountain" idea is that the transmitter encodes a finite length information sequence into a potentially infinite stream of encoded symbols; the receiver then accumulates received symbols (possibly noisy) until it finds that it has enough for successful decoding.

The first codes of this type were the LT ("Luby Transform") codes of Luby [119], in which each encoded symbol is a parity check on a randomly chosen subset of the information symbols. These were extended to the "Raptor codes" of Shokrollahi [120], in which an inner LT code is concatenated with an outer fixed-length, high-rate LDPC code. Raptor codes permit linear-time decoding and clean up error floors, with a slightly greater coding overhead than LT codes. Both types of codes work well on erasure channels, and both have been implemented for Internet applications by Digital Fountain, Inc.. Raptor codes also appear to work well over more general noisy channels, such as the AWGN channel [121].

### F. Approaching the capacity of bandwidth-limited channels

In Section V, we discussed coding for bandwidth-limited channels. Following the introduction of capacity-approaching codes, researchers turned their attention to applying these new techniques to bandwidth-limited channels. Much of the early research followed the approach of Ungerboeck's trellis-coded modulation [77] and the related work of Imai and Hirakawa on multilevel coding [78]. In two variations, turbo TCM due to Robertson and Woerz [122] and parallel concatenated TCM due to Benedetto et al. [123], Ungerboeck's set partitioning rules were applied to turbo codes with TCM constituent encoders. In another variation, Wachsmann and Huber [124] adapted the multilevel coding technique to work with turbo constituent codes. In each case, performance approaching the Shannon limit was demonstrated at spectral efficiencies $\eta > 2$ b/s/Hz with large pseudorandom interleavers.

Even earlier, a somewhat different approach had been introduced by LeGoff, Glavieux, and Berrou [125]. They employed turbo codes in combination with bit-interleaved coded modulation (BICM), a technique originally proposed by Zehavi [126] for bandwidth efficient convolutional coding on fading channels. In this arrangement, the output sequence of a turbo encoder is bit-interleaved and then Gray-mapped directly onto a signal constellation, without any attention to set partitioning or multilevel coding rules. However, because turbo codes are so powerful, this seeming neglect of efficient signal mapping design rules costs only a small fraction of a dB for most practical constellation sizes, and capacity-approaching performance can still be achieved. In more recent years, many variations of this basic scheme have appeared in the literature. A number of researchers have also investigated the use of LDPC codes in BICM systems. Because of its simplicity and the fact that coding and signal mapping can be considered separately, combining turbo or LDPC codes with BICM has become the most common capacity-approaching coding scheme for bandwidth-limited channels.

### G. Codes on graphs

The field of "codes on graphs" has been developed to provide a common conceptual foundation for all known classes of capacity-approaching codes and their iterative decoding algorithms.

Inspired partly by Gallager, Michael Tanner founded this field in a landmark paper nearly 25 years ago [88]. Tanner introduced the Tanner graph bipartite graphical model for LDPC codes, as shown in Figure 13. Tanner also generalized the parity-check constraints of LDPC codes to arbitrary linear code constraints. He observed that this model included product codes, or more generally codes constructed "recursively" from simpler component codes. He derived the generic sum-product decoding algorithm, and introduced what is now called the "min-sum" (or "max-product") algorithm. Finally, his grasp of the architectural advantages of "codes on graphs" was clear:

> The decoding by iteration of a fairly simple basic operation makes the suggested decoders naturally adapted to parallel implementation with large-scale-integrated circuit technology. Since the decoders can use soft decisions effectively, and because of their low computational complexity and parallelism can decode large blocks very quickly, these codes may well compete with current convolutional techniques in some applications.

Like Gallager's, Tanner's work was largely forgotten for many years, until Niclas Wiberg's seminal thesis [86], [87]. Wiberg based his thesis on LDPC codes, Tanner's paper, and the field of "trellis complexity of block codes," discussed in Section IV-E above.

Wiberg's most important contribution may have been to extend Tanner graphs to include state variables as well as symbol variables, as shown in Figure 16(a). A Wiberg-type graph, now called a "factor graph" [127], is still bipartite; however, in addition to symbol variables, which are external, observable, and determined *a priori*, a factor graph may include state variables, which are internal, unobservable, and introduced at will by the code designer.

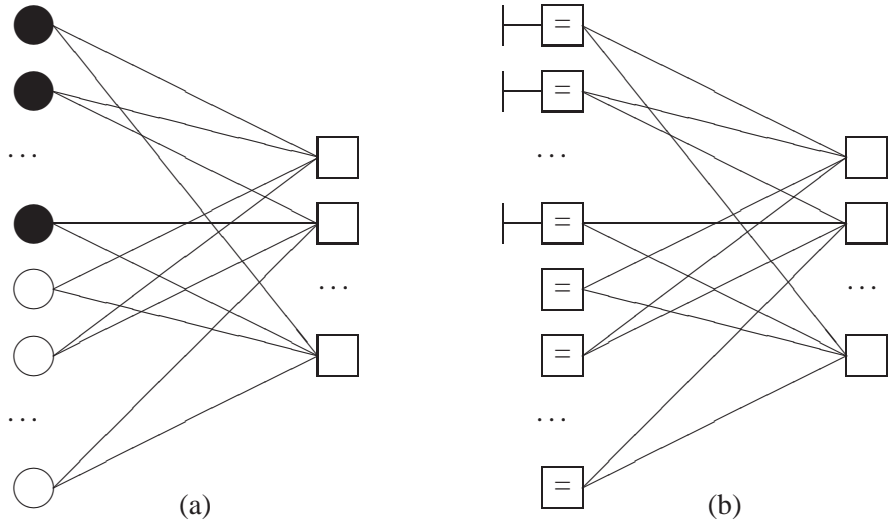

Fig. 16. (a) Generic bipartite factor graph, with symbol variables (filled circles), state variables (open circles), and constraints (squares). (b) Equivalent normal graph, with equality constraints replacing variables, and observed variables indicated by "half-edges."

Subsequently Forney proposed a refinement of factor graphs, namely "normal graphs" [128]. Figure 16(b) shows a normal graph that is equivalent to the generic factor graph of Figure 16(a). In a normal graph, state variables are associated with edges and symbol variables with "half-edges;" state nodes are replaced by repetition constraints that constrain all incident state edges to be equal, while symbol nodes are replaced by repetition constraints and symbol half-edges. This conversion thus causes no change in graph topology or complexity. Both styles of graphical realization are in current use, as are "Forney-style factor graphs."

By introducing states, Wiberg showed how turbo codes and trellis codes are related to LDPC codes. Figure 17(a) illustrates the factor graph of a conventional trellis code, where each constraint determines the possible combinations of (state, symbol, next state) that can occur. Figure 17(b) is an equivalent normal graph, with state variables represented simply by edges. Note that the graph of a trellis code has no cycles (loops).

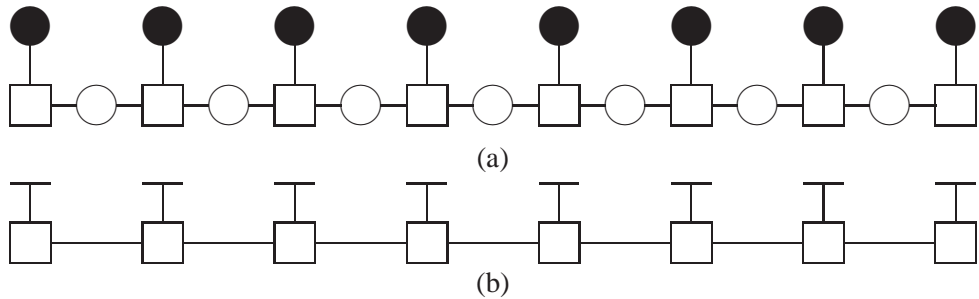

Fig. 17. (a) Factor graph of a trellis code, (b) equivalent normal graph of a trellis code.

Perhaps the key result following from the unification of trellis codes and general codes on graphs is the "cut-set bound," which we now briefly describe. If a code graph is disconnected into two components by deletion of a cut set (a minimal set of edges whose removal partitions the graph into two disconnected components), then the code constraints require a certain minimum amount of information to pass between the two components. In a trellis, this establishes a lower bound on state space size. In a general graph, it establishes a lower bound on the *product* of the sizes of the state spaces corresponding to a cut set. The cut-set bound implies that cycle-free graphs cannot have state spaces much smaller than those of conventional trellises, since cut sets in cycle-free graphs are single edges; dramatic reductions in complexity can occur only in graphs with cycles, such as the graphs of turbo and LDPC codes.

In this light turbo codes, LDPC codes, and RA codes can all be seen as codes whose graphs are made up of simple codes with linear-complexity graph realizations, connected by a long, pseudo-random interleaver $\Pi$, as shown in Figures 18, 19, and 20.

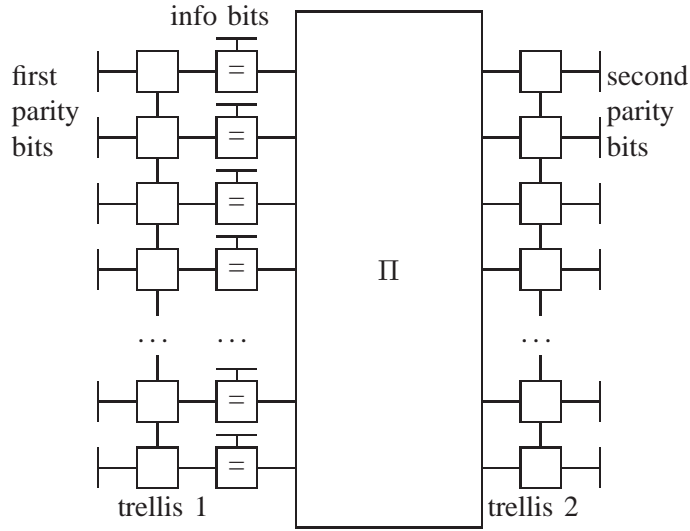

Fig. 18. Normal graph of a Berrou-type turbo code. A data sequence is encoded by two low-complexity trellis codes, in one case after interleaving by a pseudo-random permutation $\Pi$.

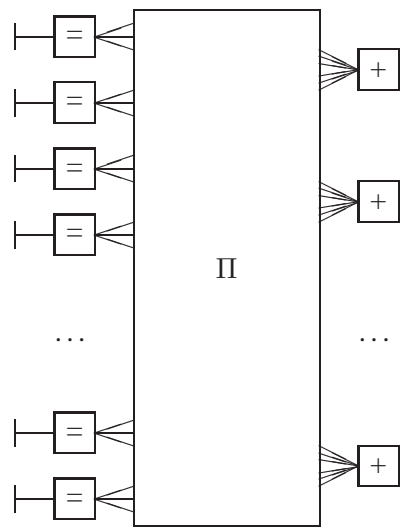

Fig. 19. Normal graph of a regular $d_\lambda = 3, d_\rho = 6$ LDPC code. Code bits satisfy single-parity-check constraints (indicated by "+"), with connections specified by a pseudo-random permutation $\Pi$.

Wiberg made equally significant conceptual contributions on the decoding side. Like Tanner, he gave clean characterizations of the min-sum and sum-product algorithms, showing that they were essentially identical except for the substitution of "min" for "sum" and "sum" for "product" (and even giving the further "semi-ring" generalization

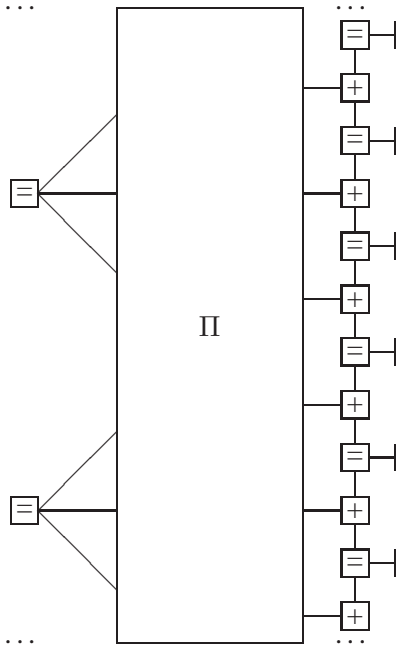

Fig. 20. Normal graph of rate-$\frac{1}{3}$ RA code. Data bits are repeated three times, permuted by a pseudo-random permutation $\Pi$, and encoded by a rate-1/1 convolutional encoder.

[129]). He showed that on cycle-free graphs they perform exact ML and APP decoding, respectively. In particular, on trellises they reduce to the Viterbi and BCJR algorithms, respectively.[15] This result strongly motivates the heuristic extension of iterative sum-product decoding to graphs with cycles. Wiberg showed that the turbo and LDPC decoding algorithms may be understood as instances of iterative sum-product decoding applied to their respective graphs. While these graphs necessarily contain cycles, the probability of short cycles is low, and consequently iterative sum-product decoding works well.

Forney [128] showed that with a normal graph representation, there is a clean separation of functions in iterative sum-product decoding:

- All computations occur at constraint nodes, not at states;
- State edges are used for internal communication (message-passing);
- Symbol edges are used for external communication (I/O).

Connections shortly began to be made to a variety of related work in various other fields, notably in [130], [131], [129], [127]. In addition to the Viterbi and BCJR algorithms for decoding trellis codes and the turbo and LDPC decoding algorithms, the following algorithms have all been shown to be special cases of the sum-product algorithm operating on appropriate graphs:

- the "belief propagation" and "belief revision" algorithms of Pearl [132], used for statistical inference on "Bayesian networks;"
- the "forward-backward" (Baum-Welch) algorithm [133], used for detection of hidden Markov models in signal processing, especially for speech recognition;
- "Junction tree" algorithms used with Markov random fields [129]; and
- Kalman filters and smoothers for general Gaussian graphs [127].

In summary, the principles of all known capacity-approaching codes and a wide variety of message-passing algorithms used not only in coding but also in computer science and signal processing can be understood within the framework of "codes on graphs."

---

[15]Indeed, the "extrinsic" APP's passed in a turbo decoder are exactly the messages produced by the sum-product algorithm.

TABLE I

APPLICATIONS OF TURBO CODES (COURTESY OF C. BERROU).

| Application | turbo code | termination | polynomials | rates |
|---|---|---|---|---|
| CCSDS (deep space) | binary, 16-state | tail bits | 23, 33, 25, 37 | 1/6, 1/4, 1/3, 1/2 |
| UMTS, CDMA2000 (3G Mobile) | binary, 8-state | tail bits | 13, 15, 17 | 1/4, 1/3, 1/2 |
| DVB-RCS (Return Channel over Satellite) | duo-binary, 8-state | circular | 15, 13 | 1/3 up to 6/7 |
| DVB-RCT (Return Channel over Terrestial) | duo-binary, 8-state | circular | 15, 13 | 1/2, 3/4 |
| Inmarsat (Aero-H) | binary, 16-state | no | 23, 35 | 1/2 |
| Eutelsat (Skyplex) | duo-binary, 8-state | circular | 15, 13 | 4/5, 6/7 |
| IEEE 802.16 (WiMAX) | duo-binary, 8-state | circular | 15, 13 | 1/2 up to 7/8 |

*Notes*:

1) "duo-binary" refers to a turbo code with rate-2/3 constituent codes;
2) "termination" refers to the method of forcing the encoder back to a known state following encoding;
3) polynomials, given in octal notation, specify encoder connections.

*H. The impact of the turbo revolution*

Even though it has been less than 15 years since the introduction of turbo codes, these codes and the related class of LDPC codes have already had a significant impact in practice. In particular, almost all digital communication and storage system standards that involve coding are being upgraded to include these new capacity-approaching techniques.

Known applications of turbo codes as of this writing are summarized in Table I. LDPC codes have been adopted for the DVB-S2 (Digital Video Broadcasting) and 10GBASE-T or IEEE 802.3an (Ethernet) standards, and are currently also being considered for the IEEE 802.16e (WiMax) and 802.11n (WiFi) standards, as well as for various storage system applications.

It is evident from this explosion of activity that capacity-approaching codes are revolutionizing the way that information is transmitted and stored.

## VII. CONCLUSIONS

It took only 50 years, but the Shannon limit is now routinely being approached within 1 dB on AWGN channels, both power-limited and bandwidth-limited. Similar gains are being achieved in other important applications, such as wireless channels and Internet (packet erasure) channels.

So is coding theory finally dead? The Shannon limit guarantees that on memoryless channels such as the AWGN channel, there is little more to be gained in terms of performance. Therefore channel coding for classical applications has certainly reached the point of diminishing returns, just as algebraic coding theory had by 1971.

However, this does not mean that research in coding will dry up, any more than research in algebraic coding theory has disappeared. There will always be a place for discipline-driven research that fills out our understanding. Research motivated by issues of performance *vs.* complexity will always be in fashion, and measures of "complexity" are sure to be redefined by future generations of technology. Coding for non-classical channels, such as multi-user channels, networks, and channels with memory, are hot areas today that seem likely to remain active for a long time. The world of coding research thus continues to be an expanding universe.

## ACKNOWLEDGMENTS

The authors would like to thank Mr. Ali Pusane for his help in the preparation of this manuscript. Comments on earlier drafts by C. Berrou, J. L. Massey, and R. Urbanke were very helpful.